\documentclass[%
 reprint,
superscriptaddress,
 amsmath,amssymb,
 aps,pra
floatfix,
]{revtex4-1}

\usepackage{graphicx}% Include figure files
\usepackage{dcolumn}% Align table columns on decimal point
\usepackage{bm}% bold math
\usepackage{tipa}
\usepackage{epsfig}
\usepackage{amsmath}   
\usepackage{amsfonts}
\usepackage{amssymb}
\usepackage{graphicx}  
\usepackage{verbatim} 
\usepackage{color}   
\usepackage{subfigure}  
\usepackage{hyperref}  
\usepackage[T1]{fontenc}
\usepackage{xcolor}
\newcommand{\isKL}{isKL}
\newcommand{\inPCA}{inPCA}

\newcommand{\dsBhat}{\ensuremath{d_{sBhat}}}
\newcommand{\btheta}{{\mbox{\boldmath$\theta$}}}
\newcommand{\bgamma}{{\mbox{\boldmath$\tilde{\theta}$}}}
\newcommand{\bthetaa}{{\mbox{\boldmath$\color{white}{\tilde{\color{black}{\theta}}}$}}}

\graphicspath{{Fig/}}
\newcommand{\Renyi}{R{\'e}nyi}
\newcommand{\phib}{\langle\Phi_{i}\rangle}

\begin{document}

\preprint{APS/123-QED}

\title{Visualizing probabilistic models in Minkowski space with intensive symmetrized Kullback-Leibler
 embedding}% Force line breaks with \\

\author{Han Kheng Teoh}
\affiliation{LASSP, Physics Department, Cornell University, Ithaca, NY 14853-2501, United States}
\author{ Katherine N. Quinn}
\affiliation{Center for the Physics of Biological Function, Department of Physics, Princeton University, Princeton NJ}
\affiliation{Initiative for Theoretical Sciences, the Graduate Center CUNY, New York NY}
\author{Jaron Kent-Dobias}
\affiliation{LASSP, Physics Department, Cornell University, Ithaca, NY 14853-2501, United States}
\author{Qingyang Xu}
\affiliation{MIT Operations Research Center, Cambridge, MA 02139, United States}
\author{James P. Sethna}
\affiliation{LASSP, Physics Department, Cornell University, Ithaca, NY 14853-2501, United States}

\date{\today}% It is always \today, today,

\begin{abstract}
We show that the predicted probability distributions for any $N$-parameter
statistical model taking the form of an exponential family can be explicitly
and analytically embedded isometrically in a $N{+}N$-dimensional Minkowski
space. That is, the model predictions can be visualized as control parameters
are varied, preserving the natural distance between probability distributions.
All pairwise distances between model instances are given by the symmetrized
Kullback-Leibler divergence. We give formulas for these intensive symmetrized Kullback Leibler (isKL) coordinate
embeddings, and illustrate the resulting visualizations with the Bernoulli (coin toss)
problem, the ideal gas, $n$ sided die, the nonlinear least squares fit, and
the Gaussian fit. We highlight how isKL can be used to determine the minimum number of parameters needed to describe probabilistic data, and conclude by visualizing the prediction space of the
two-dimensional Ising model, where we examine the manifold behavior near its
critical point. 
\end{abstract}

\maketitle

%\tableofcontents
\section{Context}
\label{sec:Context}

Many features of multiparameter models are best understood by studying the
manifold of model predictions~\cite{transtrum2015perspective}. Within this paradigm, a \textit{model manifold} is constructed, representing the space of possible model predictions. The manifold is embedded in a larger \textit{behavior space}, representing the space of all possible observables and experimental measurements.
Surprisingly, model manifolds are usually observed to be well approximated by a
relatively flat \textit{hyperribbons}, defined as objects whose successive cross sectionals are successively smaller by a roughy constant factor~\cite{TranstrumMS10, TranstrumMS11}. This has now been found in numerous nonlinear least squares models~\cite{QuinnWTS19},
and helps explain the parameter indeterminacy or `sloppiness' observed in
systems biology~\cite{GutenkunstWCBMS07}, quantum Monte Carlo~\cite{WaterfallCGBMBES06} and critical phenomena~\cite{MachtaCTS13}.  The hyperribbon geometry of the model manifold
has inspired new algorithms for nonlinear least-squares
fits~\cite{TranstrumMS10, TranstrumMS11, TranstrumSnna, TranstrumSnnb} and for
the control of complex instrumentation such as particle
accelerators~\cite{BerganBCLRS19}.

Many statistical models are not of least-squares form. For example, the Ising
model of magnetism and the $\Lambda$CDM model of the cosmic microwave
background predict the underlying statistics for experimental observation, or more generally a distribution of possible observations.  Local analysis of parameter sensitivity shows that the Ising
model~\cite{MachtaCTS13} and the $\Lambda$CDM model~\cite{QuinnCdBNS19} are
\textit{sloppy}, in the sense that they have a hierarchy of sensitivity
eigenvalues spanning many decades. These local sensitivities are quantitatively measured by
the natural distance in the space of probability distributions, the Fisher
Information Metric (FIM) \cite{amari2007methods}. 

In reference~\cite{QuinnCdBNS19} it was shown that the model manifold of probability distributions can be visualized using InPCA by embedding in a Minkowski space. For a model whose parameters $\btheta$ correspond to a probability distribution $P_\btheta (x)$ over observable data $x$, InPCA
allows visualization of the model manifold with pairwise distances between models with parameters $\btheta$ and $\bgamma$  given by the
Bhattacharyya divergence~\cite{bhattacharyya1946measure}
\begin{equation}
\label{eq:symBhat}
D^{2}_{Bhat}(P_\bthetaa,P_\bgamma) 
	= - \log\left(\sum_x \sqrt{P_\bthetaa(x) P_\bgamma(x)}\right).
\end{equation}
For the Ising and $\Lambda$CDM models, $x$ runs over spin configurations and
observed spatial CMB maps respectively. The manifold visualized with InPCA reveals its hyperribbon structure, thereby
capturing most of the model variation with only a few principal components.
The key trick in InPCA, where the limit of zero data is considered to extract an \textit{intensive} property, can be applied using a more
general class of pairwise distances given by the $f$ divergences \cite{csiszar2004information} and
in return yields a collection of intensive distance measures, expressed as a
linear combinations of the R\'enyi divergences \cite{renyi1961measures} (
details of which are provided in Appendix A).  All R\'enyi divergences locally reproduce
the FIM, so distances in behavior space reflect how sensitive the model
predictions are to shifts in the model parameters.

Here we show, for a large class of important multiparameter models, that a  different intensive embedding, built on the symmetrized Kullback-Leibler
divergence \cite{kullback1951information}
\begin{equation}
\label{eq:dsymKL}
D^{2}_{sKL}(P_{\bthetaa},P_{\bgamma}) 
	= \sum_x (P_{\bthetaa}(x)-P_{\bgamma}(x))
			      \log{\left(\frac{P_{\btheta}(x)}{P_{\bgamma}(x)}\right)}
\end{equation}
generates an explicit, analytically tractable embedding in a Minkowski space of
dimension equal to twice the number of parameters. We call this the \isKL\
embedding ({\em i}ntensive {\em s}ymmetrized {\em K}ullback-{\em L}eibler,
pronounced 'icicle'), and provide the corresponding \isKL\ coordinates in
Sec.~\ref{sec:isKLCoordinates}.  Our result is obtained for models which form
the {\em exponential families} \cite{nielsen2009statistical}:
\begin{equation}
\label{eq:exponentialFamily}
P_{\boldsymbol{\theta}}(x)=h(x)\exp\left(\sum_{i} \eta_i(\btheta) \Phi_{i}(x)-A(\boldsymbol{\theta})\right),
\end{equation}
where $h(x)$ is the base measure, $\eta_i(\btheta)$ is the $i$th natural
parameter, $\Phi_i(x)$ is the $i$th sufficient statistic, and
$A(\btheta)$ is the log partition function. Many models in
statistical mechanics form exponential families, e.g., the Boltzmann
distribution defined on most Hamiltonians. Moreover, while our method can be used to visualize the manifolds of probabilistic models described by exponential families, we explain in Sec.~\ref{sec:Examples} how we can use this method to determine the minimum number of parameters needed to describe probabilistic data.

\section{Curse of Dimensionality}
\label{sec:curseofdim}

Large data sets and multiparameter probabilistic models of large systems both suffer from the curse of dimensionality~\cite{Kriegel2009ClusteringHD}: as the dimension of the system increases, it becomes more difficult to establish meaningful relationships between points as the distance measure becomes saturated.
This effect obscures meaningful features within the data set and renders
contrast in distances between different data points
nonexistent~\cite{curseofdimensionality1999}. 

Intensive embeddings like \inPCA\ and \isKL\ break the curse of dimensionality
for probabilistic models, allowing for low-dimensional projections of model
manifolds in a suitable Minkowski space~\cite{QuinnCdBNS19}. Big data
applications have attempted to resolve this dimensionality issue by embedding
the manifold in a curved
space~\cite{wilson2010spherical,boguna2010sustaining,nickel2017poincare} or in
an Euclidean space with an alternative distance
measure~\cite{tenenbaum2000global,belkin2003laplacian,maaten2008visualizing,moon2019visualizing},
which can yield lower dimensional projections that capture dominant components
of the variation in the data set. For example, reference~\cite{moon2019visualizing} makes use of the extensive%
  \footnote{The potential distance between $N$ replicated system is 
  \begin{align*}
  D^{2}_{pot}(p^{N},q^{N})&=\sum_{i}\log^{2}(p_{i}^{N}/q_{i}^{N})\\
			  &=N^{2}\sum_{i}\log^{2}(p_{i}/q_{i})\\
			  &=N^{2}D^{2}_{pot}(p,q).
   \end{align*}
Thus the potential distance per replica scales linearly with $N$.} and non-isometric
  \footnote{
The potential distance between two distributions $p$ and $q$ , 
$D^{2}_{pot}=\sum_{i}\log^{2}(p_{i}/q_{i})$ can be shown to disagree with the FIM locally as follows:
\begin{align*}
D^{2}_{pot}(\theta+\delta\theta,\theta)&=\sum_{i}\log^{2}(p_{i}(\theta+\delta\theta)/p_{i}(\theta)) \\
	&\approx\sum_{i}\log^{2}{\bigg(1+\frac{\partial_{\alpha}p_{i}(\theta)}{p_{i}(\theta)}\delta\theta^{\alpha}\bigg)}\\
&\approx\sum_{i}\bigg(\frac{\partial_{\alpha}p_{i}(\theta)}{p_{i}(\theta)}\bigg)\bigg(\frac{\partial_{\beta}p_{i}(\theta)}{p_{i}(\theta)}\bigg)\delta\theta^{\alpha}\delta\theta^{\beta} \\
	&=\sum_{i}\partial_{\alpha}\log{p_{i}(\theta)}\partial_{\beta}\log{p_{i}(\theta)}\delta\theta^{\alpha}\delta\theta^{\beta}.
\end{align*}
This is different from the FIM that takes the form $ I_{\alpha\beta}(\theta)=\sum_{i}p_{i}(\theta)\partial_{\alpha}\log{p_{i}(\theta)}\partial_{\beta}\log{p_{i}(\theta)}$ } \textit{potential distance} in generating useful visualizations of large data sets for
biological data in Euclidean space.  Our methods suggest an alternative approach. 

\begin{figure*}
    \centering
    \includegraphics[scale=0.4]{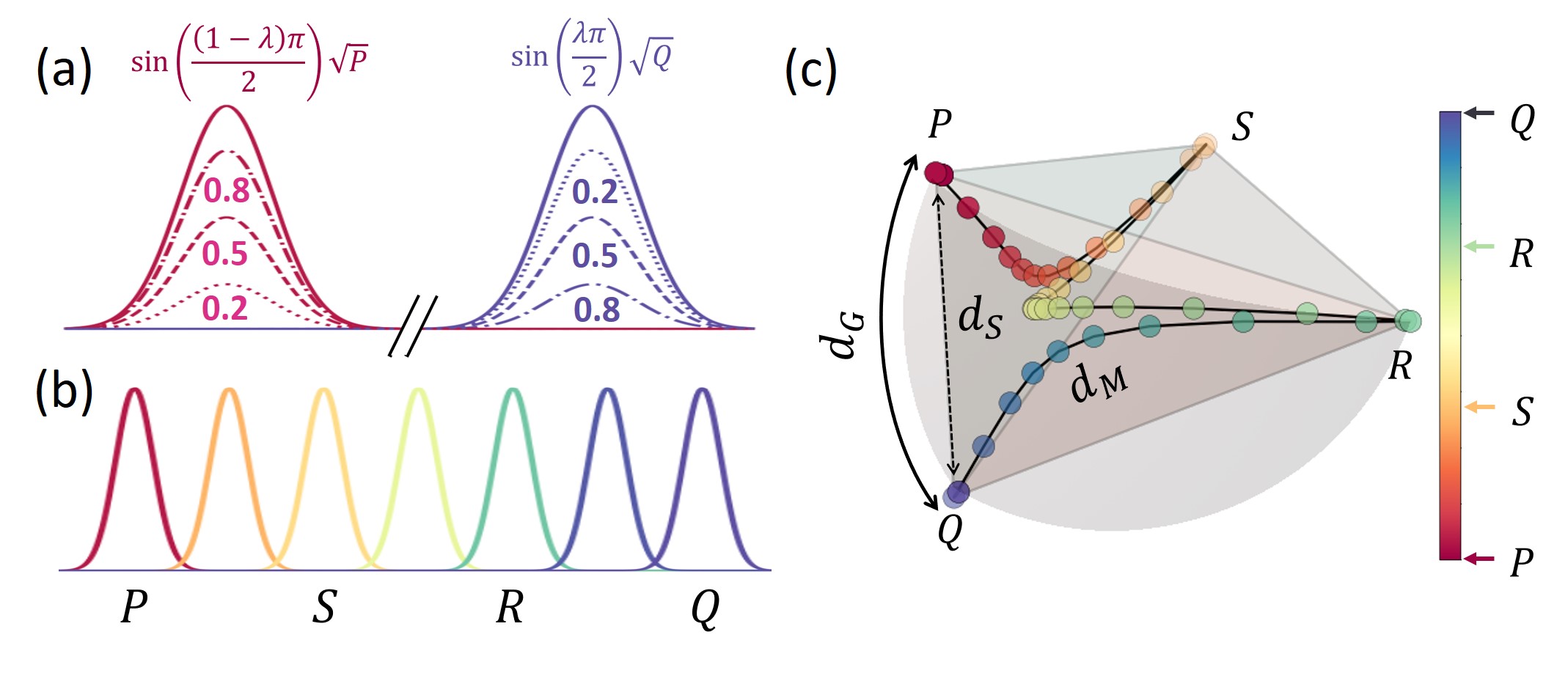}
    \caption{
   (a)  The geodesic path (with path length $d_G$) between two probability distributions \textit{P} and \textit{Q} is given by an interpolation: $  \sqrt{P^{*}_{\lambda}(x)}=\frac{\sin\left({\frac{(1-\lambda)d_{G}}{2}}\right)}{\sin\left({\frac{d_{G}}{2}}\right)}\sqrt{P(x)}+\frac{\sin\left({\frac{\lambda d_{G}}{2}}\right)}{\sin\left({\frac{d_{G}}{2}}\right)}\sqrt{Q(x)}$.
   This equals $\sin\left(\frac{(1-\lambda)\pi}{2}\right) \sqrt{P(x)} + \sin\left(\frac{\lambda\pi}{2}\right) \sqrt{Q(x)}$ in the limit when $P$ and $Q$ are orthogonal.  As $0\le\lambda\le 1$, the interpolation remains positive and normalized.
   The length of this path under the Fisher Information Metric (FIM)
   equals the arc length of the great circle, which is 
   $d_{G}(P,Q) = 2 \arccos{\sum_x \sqrt{P(x)} \sqrt{Q(x)}}$. 
   b) The shortest path through the model manifold, with path length given by $d_M$ between two Gaussian distribution with fixed $\sigma$ is given by sliding the Gaussian $\mu_{P}$ to $\mu_{Q}$, $d_{M}=\sigma^{-1}|\mu_{P}-\mu_{Q}|$. c)  The global pairwise distance. $d_S$, between distributions as compared to $d_G$ and $d_M$. The pairwise distance $d_S$ is determined by the Euclidean distance between points, and is represented here as a straight line from \textit{P} to \textit{Q}. The octant of 
   the sphere schematically represents the space of all possible probability
   distributions (due to the normalized, non-negative nature of distributions, discussed further in Sec. II), and so the great circle path $d_G$ is the arc length distance from \textit{P} to \textit{Q}. 
   The manifold path length $d_M$ is the minimum distance
   between the two distributions when one considers the path through the model manifold of a complex, nonlinear model. When $d_{M}\gg d_{G}$, the path must curl around in multiple dimensions to fit inside the sphere;
   mutually orthogonal distributions (as in (b)) will form a hyper-tetrahedron
   inside the model manifold. Note that (c) represents a 3D projection of a much higher dimensional space.}
    \label{fig:curse_of_dim}
\end{figure*}

To prove the utility of embedding probability distributions in a Minkowski space,  we consider discrete probability distributions, $\sum_{x}P(x)=1$ for simplicity. We first introduce three type of distances, namely (1) a geodesic distance $d_G$ between two distributions, defined as the shortest path through the space of all possible probability distributions. Because probability distributions are normalized and non-negative, they can be interpreted as unit vectors $y(x) = \sqrt{P(x)}$  in a high-dimensional space, thus forming a high-dimensional sphere. Thus, the path between two distributions would be the great-circle distance between them, (2)  the shortest path-length distance through the model manifold, which we call the manifold distance $d_M$ and (3) the pairwise straight-line distance $d_S$ in the embedding space, with a metric which here will be a particular divergence measure. These distances are illustrated in Fig.~\ref{fig:curse_of_dim}(c). 
First, note that the path-length distance for $d_G$ and $d_M$ is computed by integrating the Fisher Information Metric (FIM) along said path:
\begin{equation}
    \label{eq:FIM}
    I_{\alpha,\beta}(\btheta)=-\bigg\langle \frac{\partial^{2}\log {P(x)}}{\partial\theta_{\alpha}\partial\theta_{\beta}}\bigg\rangle_{x}
\end{equation}
giving
\begin{equation}
\label{eq:StraightPathBound}
d(P,Q)=\int\sqrt{\sum_{x}\frac{1}{P^{*}_{\lambda}(x)}\bigg(\frac{d P^{*}_{\lambda}(x)}{d\lambda}\bigg)^{2}}d\lambda,  
\end{equation}
where $\lambda$ parametrizes the path between $P(x)$ and $Q(x)$. Upon letting $P^{*}_{\lambda}(x)=y^{2}_{\lambda}(x)$, Eq.~\eqref{eq:StraightPathBound} simplifies to 
\begin{equation}
d(P,Q)=\int 2\sqrt{\sum_{x} \left(\frac{dy_{\lambda}(x)}{d\lambda}\right)^{2}}d\lambda,     
\end{equation}
representing the familiar path length in Euclidean space. The requirement $\sum_{x}P^{*}_{\lambda}(x)=\sum_{x}y^{2}_{\lambda}(x)=1$  restricts the path to lie on a sphere, thus Eq. \eqref{eq:StraightPathBound} yields the arc length of a great circle connecting the
two distributions \cite{wootters1981statistical},  
\begin{equation}
\label{eq:greatcircledistance}
    d_{G}(P,Q) =2 \arccos{\sum_x \sqrt{P(x)} \sqrt{Q(x)}}.
\end{equation}
Alternatively, one could perform a variational calculation on Eq.~\eqref{eq:StraightPathBound} to find the shortest path connecting $P(x)$ and $Q(x)$ and its length. This has been worked out in \cite{ito2018stochastic} and the path is given by 
\begin{equation}
    \sqrt{P^{*}_{\lambda}(x)}=\frac{\sin\left({\frac{(1-\lambda)d_{G}}{2}}\right)}{\sin\left({\frac{d_{G}}{2}}\right)}\sqrt{P(x)}+\frac{\sin\left({\frac{\lambda d_{G}}{2}}\right)}{\sin\left({\frac{d_{G}}{2}}\right)}\sqrt{Q(x)},
\end{equation}
where $\lambda\in[0,1]$ and $d_{G}$ is given by Eq.~\eqref{eq:greatcircledistance}.Thus the geodesic path between two probability distributions $P$ and $Q$ is a linear interpolation between $\sqrt{P}$ and $\sqrt{Q}$, renormalized to unit length.

When considering a more specific path through a model manifold generated by a specific physical model (\textit{e.g.} the Ising model), Eq.~\eqref{eq:StraightPathBound} no longer reduces to such a simple form. Instead, we obtain a more complicated expression, which represents a path through the manifold, $d_M$. However, since the manifold is confined to the surface of the hypersphere, it is bounded from below by the geodesic distance $d_G \leq d_M$.  To illustrate the difference between $d_G$ and $d_M$, consider the example illustrated in Fig. \ref{fig:curse_of_dim}: If \textit{P} and \textit{Q}
are non-overlapping Gaussians of mean $\mu_P$ and $\mu_Q$, the geodesic path $d_M$ along the
model manifold of Gaussians of fixed width $\sigma$ is given by sliding the 
Gaussian from $\mu_P$ to $\mu_Q$, while the shortest path in the space
of all probability distributions $d_{G}$ is given by shrinking $P$ and growing
$Q$ in place (see Fig.~\ref{fig:curse_of_dim}(a) and (b)).

A key point is that for any embedding that takes general families
of probability distributions isometrically into a Euclidean space,
{\em the straight line distance $d_S$ is constrained by the diameter of the hypersphere containing the probability distributions} (Eq. \ref{eq:greatcircledistance}). 
To further illustrate the differences between the three type of distances, we embedded simple Gaussians with a fixed variance on a hypersphere by using the Hellinger divergence  $d^{2}_{Hel}=\sum_{x}(\sqrt{P(x)}\cdot\sqrt{Q(x)})^{2}$ as the straight line distance $d_S$.  Fig.~\ref{fig:curse_of_dim}(c) depicts the three-dimensional projection of the Hellinger embedding. Here, the octant represents the space of all possible probability distribution schematically. In our simple example, if $\mu_P$ and $\mu_Q$ are many standard deviations
apart, the path length distance $d_{M}$ between them on the fixed-Gaussian model manifold
has length
\begin{equation}
d_{M} =\int_{\mu_{P}}^{\mu_{Q}} \frac{d\mu}{\sigma}=   \frac{|\mu_P-\mu_Q|}{\sigma}.
\end{equation}
When $d_{M} \gg d_{G}$, the path must curl around to fit inside the sphere of
radius $2$. Thus, low-dimensional projection will at best show a crumpled
tangle that usually rapidly escapes into higher, undisplayed dimensions. In other words, a useful low-dimensional projection should be able to take any set of $M$ probability distributions and project them in a way that maintains their distinguishability. More generally, the low-dimensional projection should take $M$ probability distributions with mutual near zero overlap and keep them separated by at least some minimum embedding-space distance $\Delta$, presumably comparable to $\pi$. The minimum embedding
dimension for such a set of points is given by the densest packing of spheres
of diameter $\Delta$ into a sphere of diameter $d_{G}$ in $D$ dimensions. For
the Hellinger embedding or whenever $\Delta \sim 1$, one needs $\sim M-1$
projection directions to accurately visualize $M$ mutually distinguishable predictions.

Note that in an Euclidean space, the global pairwise distance $d_{S}$ is always
smaller than geodesic path through the hypersphere (great circle length) $d_{G}$ (bounded by $\pi$, Eq.~\eqref{eq:greatcircledistance}). The geodesic distance $d_G$ sets a lower bound on the manifold path length, $d_M$, since the manifold is confined to the surface of the hypersphere. We shall illustrate many
times in the rest of this manuscript that this bound no longer holds when one considers embeddings
in Minkowski space. These Minkowski space embeddings can be constructed by defining a pairwise distance between probability distributions $d_{S}$ that violates the triangle inequality which in turn breaks the curse of dimensionality as noted in ~\cite{QuinnCdBNS19}.  For example, Fig.~4 in reference~\cite{QuinnCdBNS19}
shows the InPCA model manifold for the coin-flip problem 
(different from the \isKL\ embedding in section~\ref{subsec:cointoss}).
The straight-line distance between the two end-points (all heads and all
tails) in Minkowski space goes to infinity, but the model manifold 
hugs a light cone, and the embedding distances from either endpoint to
a fair coin is finite. We have shown here how the curse of dimensionality manifests in the Euclidean space of probability distributions. To circumvent this problem, we instead consider embeddings in a Minkowski space, and develop our isKL method in the following section.

\section{{isKL} Coordinates} 
\label{sec:isKLCoordinates}

In this section we derive the \isKL\ coordinates for a general exponential
family, giving an explicit isometric embedding for probability distributions in a Minkowski space.
Our embedding space is similar to Minkowski space but not identical to it, in that it has $N$ space-like coordinates (positive metric elements) with $N$ corresponding time-like coordinates (negative metric elements), forming an $N+N$ dimensional space. We shall generate two coordinates
$\mathcal{S}_i(\btheta)$ and $\mathcal{T}_i(\btheta)$ for each
natural parameter $\eta(\btheta)$, one
space-like (with positive squared distance) and one time-like (with
negative squared distance), such that 
\begin{equation}
\label{eq:isKLsum}
\begin{split}
D^{2}_{sKL}(P_\bthetaa,P_\bgamma) 
	&= \sum_i (\mathcal{S}_i(\btheta) - \mathcal{S}_i(\bgamma))^2\\&
	- \sum_{i}(\mathcal{T}_i(\btheta) - \mathcal{T}_i(\bgamma))^2.
\end{split}
\end{equation}
where $P_{\bthetaa}$ and $P_{\bgamma}$ are two probability distributions produced by the model for parameters evaluated at $\btheta$ and $\bgamma$. The squared term with a positive sign is thus a space-like coordinate, and the
term with a negative sign is the corresponding time-like coordinate.  Since the
symmetrized Kullback-Leibler distance is nonnegative, no pair of points can be
time-like separated. The length of the model manifold projection along the time-like coordinates will typically be smaller than the length of its projection along the space-like coordinates. However, the time-like coordinates are both physical
and important, as we shall illustrate in particular using the 2D Ising model. 

The symmetrized K-L divergence ($D_{sKL}^2$) from Eq.~\eqref{eq:dsymKL}, evaluated for the exponential families considered in this manuscript (shown in Eq.\eqref{eq:exponentialFamily}), reduces to:
\begin{equation}
\begin{aligned}
 D^{2}_{sKL}(P_\bthetaa,P_\bgamma) = \sum_{i}
		\left(\eta_{i}(\btheta) - \eta_{i}(\bgamma)\right)
		\left(\langle \Phi_{i} \rangle_\btheta -
		      \langle \Phi_{i} \rangle_\bgamma\right).
\end{aligned}
\end{equation}
Now, notice that we can rearrange the terms in the above equation, we obtain:
\begin{equation}
\begin{aligned}
&\left(\eta_{i}(\btheta) -\eta_{i}(\bgamma)\right)
\left(\langle \Phi_{i} \rangle_\btheta -
 \langle \Phi_{i} \rangle_\bgamma\right)\\
    &=(1/4) \left(
	[\eta_{i}(\btheta) + \langle \Phi_{i} \rangle_\btheta] 
	- [\eta_{i}(\bgamma) + \langle \Phi_{i} \rangle_\bgamma] \right)^2\\
      &\quad -(1/4)\left(
	[\eta_{i}(\btheta) - \langle \Phi_{i} \rangle_\btheta] 
	- [\eta_{i}(\bgamma) - \langle \Phi_{i} \rangle_\bgamma] \right)^2\\
     &= (\mathcal{S}_{i}(\btheta) - \mathcal{S}_{i}(\bgamma))^2
	- (\mathcal{T}_{i}(\btheta) - \mathcal{T}_{i}(\bgamma))^2,
\end{aligned}
\end{equation} 
with the two Minkowski coordinates for the $i$-th statistic, determined from model parameters from Eq.~\eqref{eq:exponentialFamily}:
\begin{equation} 
\begin{aligned}
\label{eq:symKLcoords}
     \mathcal{S}_{i}(\btheta) &=    
	(1/2)\big(\eta_{i}(\btheta) + \langle \Phi_{i} \rangle_\btheta\big) \\
     \mathcal{T}_{i}(\btheta) &=    
	(1/2)\big(\eta_{i}(\btheta) - \langle \Phi_{i} \rangle_\btheta\big)
\end{aligned}
\end{equation} 
now summing to $D^{2}_{sKL}(P_\bthetaa,P_\bgamma)$.  The terms quadratic in the parameters and quadratic in the expectation values all cancel, and the cross terms give the contribution of statistic (defined from model parameters in Eq.~\eqref{eq:exponentialFamily}). From Eq.~\eqref{eq:symKLcoords}, the spacelike coordinate is indeed greater than the timelike coordinate for each parameter,  $(\mathcal{S}_{i}(\btheta)-\mathcal{S}_{i}(\bgamma))^{2}\geq(\mathcal{T}_{i}(\btheta)-\mathcal{T}_{i}(\bgamma))^{2}$. This is our main result.

\section{Families of embeddings: isometries of Minkowski space}
\label{sec:isKLFamilies}

The coordinates produced from isKL represent projections. Just as 
any rotation or translation of an object isometrically embedded in
Euclidean space forms another isometric embedding, so also there is a family of
\isKL\ embeddings formed by the isometries of Minkowski space. Translating the
coordinates can be used to center the sampled points of the model manifold;
certain Lorentz boosts can be valuable in minimizing the total squared length of the
coordinates (and hence reducing the importance of the time-like coordinates).
The rotational isometries within the space-like and time-like subspaces can then be
used to focus attention on the directions of the model manifold that show the
largest variations. 

As a first step in considering the effects of these isometries, 
let us consider other embeddings, similar to Eq.~\eqref{eq:symKLcoords},
that also preserve pairwise distances. Clearly one can add a constant
$C_{i}^\pm$ to each coordinate (translations in Minkowski space). One also
notes that the two terms $\eta_{i}(\btheta)$ and $\langle\Phi_{i}\rangle_{\btheta}$ being subtracted may have different
units. This can be fixed by rescaling these two terms up and down by a scale
factor $\lambda_{i}$ with units $\sqrt{[\langle\Phi_{i}\rangle_{\btheta}]/[\eta_{i}(\btheta)]}$:
\begin{equation} 
\begin{aligned}
\label{eq:symKLcoordsLambda}
\mathcal{S}_{i}(\btheta) &=    
    (1/2)(\lambda_{i} \eta_{i}(\btheta) 
    + (1/\lambda_{i}) \langle \Phi_{i} \rangle_\btheta + C_{i}^+) \\
\mathcal{T}_{i}(\btheta) &=    
    (1/2)(\lambda_{i} \eta_{i}(\btheta) - (1/\lambda_{i}) \langle \Phi_{i} \rangle_\btheta + C_{i}^-),
\end{aligned}
\end{equation} 
with different rescaling parameter $\lambda_{i}$ and shifts $C^\pm_{i}$ 
for each pair of coordinates. 

We can view Eq.~\eqref{eq:symKLcoordsLambda} as a composition of two
transformations -- a translation and a rescaling. The translation is of course
one of our isometries. The average of $\Phi_{i}$ over
parameters $\btheta$ is written as $\langle \Phi_{i}
\rangle_{\btheta}=\langle\Phi_{i}\rangle$ in the subsequent discussion for brevity.
Rescaling by $\lambda_{i}$ corresponds to a
Lorentz boost $\mathcal{T}' = \gamma (\mathcal{T}-v\mathcal{S})$, $\mathcal{S}' = \gamma (\mathcal{S}-v\mathcal{T})$ of our time-like and
space-like coordinates, where $\gamma =
1/\sqrt{1-v^2}$:
\begin{equation}
\begin{aligned}
\mathcal{T}' &= (1/2) \gamma \left((\eta_{i}(\btheta)+\langle\Phi_{i}\rangle) 
				- v (\eta_{i}(\btheta)-\langle\Phi_{i}\rangle)\right)\\
   &= (1/2) \left( \gamma (1-v) \eta_{i}(\btheta)
			- \gamma (1+v) \langle \Phi_{i} \rangle\right)\\
   &= (1/2) \left( \lambda_{i} \eta_{i}(\btheta) 
			- (1/\lambda_{i}) \langle \Phi_{i} \rangle\right),\\
\mathcal{S}' &= (1/2) \gamma \left((\eta_{i}(\btheta)-\langle\Phi_{i}\rangle) 
				- v (\eta_{i}(\btheta)+\langle\Phi_{i}\rangle)\right)\\
   &= (1/2) \left( \gamma (1-v) \eta_{i}(\btheta) 
			+ \gamma (1+v) \langle \Phi_{i} \rangle\right)\\
   &= (1/2) \left( \lambda_{i}\eta_{i}( \btheta) 
			+ (1/\lambda_{i}) \langle \Phi_{i} \rangle\right).
\end{aligned}
\end{equation}

A natural criterion for a good projection of the model manifold would be one
which minimizes the sum of squares of the coordinates. In Euclidean space,
this just translates the manifold so that its center of mass sits at the
origin. Indeed, using $C_{i}^+$ and $C_{i}^-$ to shift our two coordinates to 
their centers of mass corresponds nicely to shifting the sampled parameters
$\eta_{i}(\btheta) \to \eta_{i}(\btheta) - \overline{\eta_{i}(\btheta)}$ and resulting means
$\langle\Phi_{i}\rangle-\overline{\langle\Phi_{i}\rangle}$ to their 
respective centers of mass. Now, presuming for simplicity that the 
data is centered, let us examine the sum of the squares of our two coordinates
$\mathcal{S}_{i}$ and $\mathcal{T}_{i}$, 
\begin{equation}
(\mathcal{S}_{i}(\btheta))^{2}+(\mathcal{T}_{i}(\btheta))^{2}=\frac{1}{2}\bigg(\lambda_{i}^{2}\eta_{i}^{2}(\btheta)+\frac{1}{\lambda_{i}^{2}}\langle\Phi_{i}\rangle^{2}\bigg)
\end{equation}
To get a good point of view in Minkowski space, we seek to minimize the sum of
squares of the coordinates by optimizing $\lambda_{i}$. This yields
$\lambda_{i}^{4}=\langle \Phi_{i}\rangle^{2}/\eta_{i}^{2}(\btheta)$. As
the parameters are shifted with respect to their centers of mass, we can recast
$\lambda_{i}=(\text{Var}(\langle\Phi_{i}\rangle)/\text{Var}(\eta_{i}))^{1/4}$,
where the variance is averaged over the ensemble of parameters and the mean
$\langle \Phi_{i}\rangle$ is taken at a fixed parameter $\btheta$. It turns out our isKL embedding has a close connection to Principal Component Analysis (PCA) and Multidimensional Scaling (MDS) techniques. We refer interested readers to Appendix~\ref{sec:connection} for an in-depth discussion.

\section{Examples}
\label{sec:Examples}

\begin{figure*}[htbp]
   \begin{center}
   \includegraphics[scale=0.5]{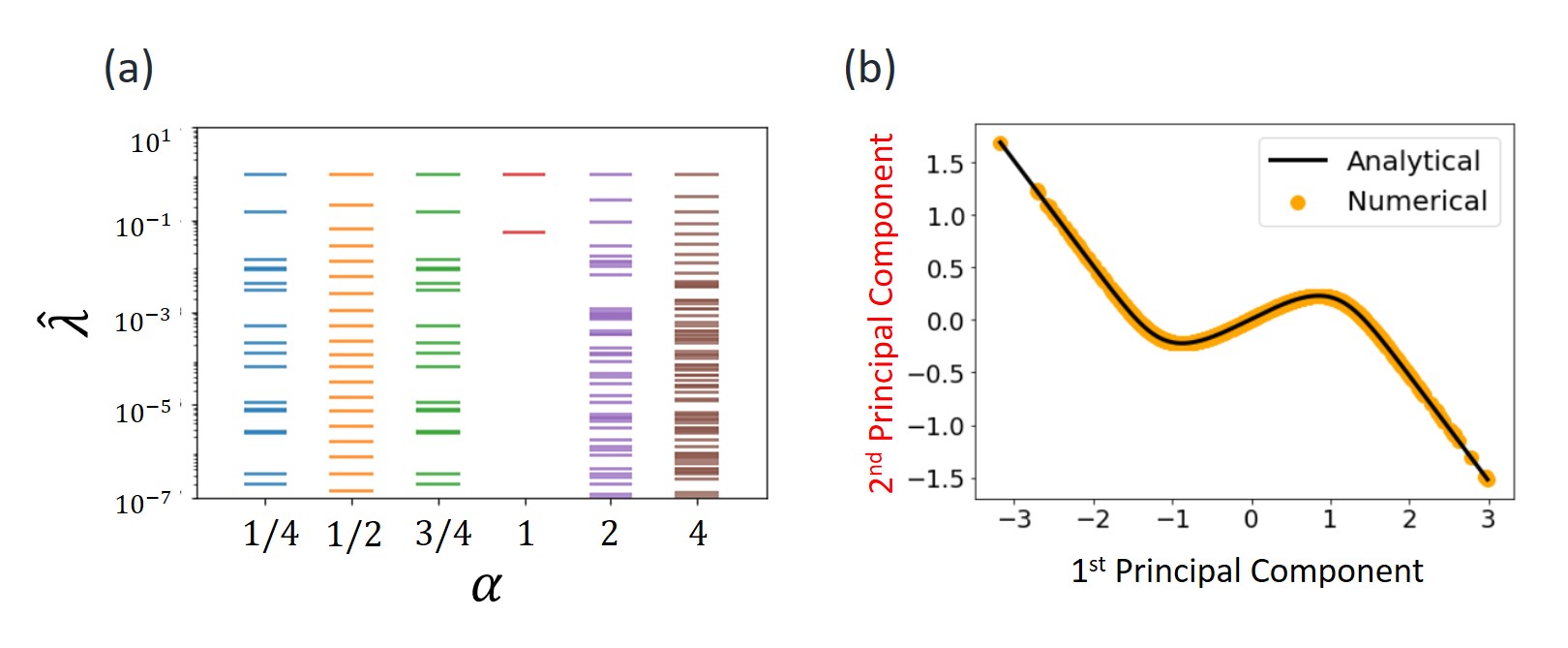}%
   \end{center}
   \caption{\label{fig:Renyichoiceformodels}
   (a)~{\bf Squared principal length of intensive embedding with different symmetrized R\'enyi choices for the coin toss manifold.} 
  $\alpha=1/2$ corresponds to Bhattacharyya divergence and $\alpha\rightarrow1$ leads to the symmetrized Kullback-Leibler divergence (sKL). Throughout the exponential family models considered in subsequent sections, sKL provides the lowest embedding dimension while other R\'enyi choices gives an embedding into which the manifold projection widths decrease geometrically over several decades. This implies the sloppiness of the embedding is influenced by the choice of divergence used. (b)  {\bf Model manifold for the Bernoulli (coin toss) Problem }~is visualized with our isKL embedding. The analytical calculation matches well with the numerical result returned from Multidimensional Scaling (MDS). }
  \label{fig:cointoss}
\end{figure*}
To demonstrate how isKL embeddings optimize the total squared distance of
coordinates to produce a good visualization, we consider several probabilistic
models that form exponential families: the Bernoulli (coin toss) problem~\ref{subsec:cointoss}, the ideal gas
model~\ref{sec:piston}, the $n$-sided die~\ref{sec:dice}, the nonlinear least square problem~\ref{subsec:NonlinearLeastSquares}, Gaussian fits to
data\ref{sec:gaussfit}, and the two dimensional Ising model~\ref{sec:ising}. We will be using $T^{\pm}_{i}(\btheta)$ to denote spacelike $\mathcal{S}_{i}(\btheta)$ and timelike $\mathcal{T}_{i}(\btheta)$ coordinates respectively for subsequent discussion for brevity. 

Before diving into the examples, it is worth highlighting that the finite
embedding dimension for exponential families appears to be a unique feature of
$D^{2}_{sKL}$. As $D^{2}_{sKL}$ is part of a family of intensive distance
measures known as the symmetrized R\'enyi divergence,  
\begin{equation}
\begin{split}
    D^{2}_{\alpha}(P,Q)=&\frac{1}{\alpha-1}\bigg(\sum_{x}\log{P(x)^{\alpha}Q(x)^{1-\alpha}}\\&+\sum_x\log{Q(x)^{\alpha}P(x)^{1-\alpha}}\bigg)
\end{split}
\end{equation} 
with $\alpha\rightarrow1$, we embed the coin toss manifold with
other symmetrized R\'enyi divergences by varying $\alpha$ to illustrate this uniqueness. As shown in Fig.
\ref{fig:cointoss} (a),  the embedding is sloppy for all $\alpha$
(geometrically decreasing manifold widths that span several decades) but only
for $\alpha=1$ does it truncate after two dimensions. This exact truncation is true for all the
probabilistic models considered in this paper. This also serves to illustrate while the symmetrized R\'enyi divergences locally reproduce the FIM that describes the local structure of a model manifold, they have a varying degree of performance in utilizing the number of dimensions to embed a model manifold isometrically. Therefore, we can reduce the embedding dimension significantly by choosing an optimal divergence. In principle, we could perform
experiments or simulations without knowing the number of parameters the
exponential family distribution needs to describe the behaviour. If the isKL
embedding gives a cutoff after $N+N$ dimensions it suggests that a hidden
$N$-parameter exponential family describes the experiment.

\subsection{Bernoulli Problem}
\label{subsec:cointoss}

The Bernoulli problem or coin tossing experiment is one of the simplest
probabilistic models. As a function of the fairness parameter $p$, the result
$x\in\{0,1\}$ of a coin toss is distributed by $P(x|p)=p^x(1-p)^{1-x}$. This
probability distribution can be written in the form of an exponential family
with $\eta(p)=\log(p/(1-p))$, $\Phi(x)=x$, $h(x)=1$, and
$A(\theta)=\log(1-e^{\theta})$.  The FIM for this model is given by
\begin{equation}
(ds)^{2}=\frac{(dp)^{2}}{p(1-p)}
\end{equation}

{\em Known embeddings:} By defining $p=\sin^{2}\theta$, we have $ds=2d\theta$. This produces a one
dimensional embedding onto a Hellinger quarter circle of radius 2 with
$\theta\in[0,\pi/2]$. Upon taking the limit of zero data, the Hellinger
distance transforms into the Bhattacharyya divergence. It is known that with the Bhattacharyya divergence, the coin toss manifold is embedded into a Minkowski space ~\cite{QuinnCdBNS19}. This embedding is illustrated in Fig~\ref{fig:Renyichoiceformodels}(a) with $\alpha=1/2$. We worked out the analytical expression for each projection coordinate in Appendix \ref{sec:coin_sol}.  Our analytical calculation suggests that the embedding is at least high dimensional. We would presume the inPCA embedding does not truncate and continue to have smaller and smaller amount of variation out to an infinite number of dimensions. 

With isKL embedding, the
coin toss manifold can be isometrically embedded into (1+1) dimensions. As $\langle\Phi \rangle=p$, its pairwise distance is given by 
 \begin{equation}
     D^{2}_{sKL}(p,q)=(p-q)\log\frac{p(1-q)}{q(1-p)}.
 \label{eq:dsklcoin}
 \end{equation}
Here, we will illustrate the utility of  Eq.~\eqref{eq:symKLcoords} in obtaining the analytical expression for each embedding coordinate. With the Jeffrey's prior as the sampling measure,  the centers of mass are $\overline{\eta}=0$ and $\overline{\langle \Phi\rangle}=1/2$ respectively. Furthermore, $\text{Var}(\eta)=\pi^{2}$ and $\text{Var}(\langle\Phi\rangle)=1/8$  we have $\lambda=(\text{Var}(\langle\Phi\rangle)/\text{Var}(\eta))^{1/4}=(2^{3/4}\sqrt{\pi})^{-1}$. With these, the spacelike and timelike $T^{\pm}(p)$ projection coordinates are determined to be   
\begin{equation}
\begin{split}
     T^{\pm}(p)&=\frac{1}{2}\bigg(\lambda (\eta-\overline{\eta})\pm \frac{1}{\lambda} \big(\Phi-\overline{\langle\Phi\rangle}\big)\bigg)\\
     &=\frac{1}{2^{7/4}\sqrt{\pi}}\log{\bigg(\frac{p}{1-p}\bigg)}\pm \frac{\sqrt{\pi}}{2^{1/4}} \bigg(p-\frac{1}{2}\bigg)
\end{split}
    \label{eq:coinflip_skl}
\end{equation}
Fig. \ref{fig:Renyichoiceformodels}(b) shows the coin toss manifold. 
\subsection{Ideal gas}
\label{sec:piston}
\begin{figure}
    \centering
    \includegraphics[scale=0.5]{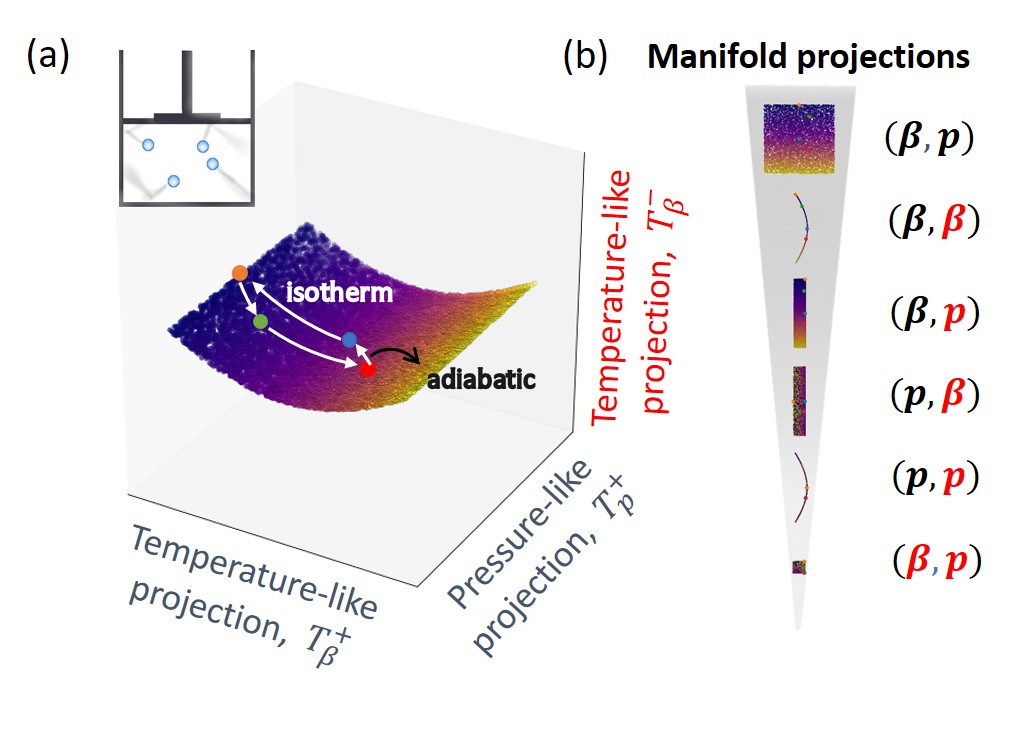}
    \caption{{\bf Model manifold for the ideal gas -}~The flat ideal gas manifold is embedded into a (2+2) dimensional Minkowski space. The manifold is 'rolled' twice in the four dimensional space, giving it a torus appearance in Minkowski space. a) The three dimensional projection of the ideal gas manifold is colored based on the inverse temperature $\beta$ and the Carnot cycle is illustrated on the manifold. b) The manifold projections are depicted in a descending order based on the manifold widths along the spacelike/timelike components. The spacelike directions are color-coded in black while the timelike directions are color-coded in red. The analytical expression for each projection is given by Eq.~\eqref{eq:coord_ideal}. The torus appearance in Minkowski space can be deduced from the curves in $(T^{+}_{\beta},T^{-}_{\beta})$ and $(T^{+}_{p},T^{-}_{p})$ coordinate pairs, both of which have the form of  $(T^{+}_{k}-C^{+}_{k})^{2}-(T^{-}_{k}-C^{-}_{k})^{2}=r^{2}$ for some constants $C^{\pm}_{k}$ and $r$. }
    \label{fig:gas}
\end{figure}
The ideal gas is a model of non-interacting particles. At pressure $P$ and temperature $\beta^{-1}$, the probability that $N$ particles will be found in a configuration with momenta $\mathbb P$, positions $\mathbb Q$, and container volume $V$ is
\begin{equation}
p(\mathbb{P},\mathbb{Q},V|P,\beta)=Z^{-1}(P,\beta)\exp{(-\beta \mathbb{P}^{2}/2m-\beta PV)},
\end{equation}
where the partition function $Z(P,\beta)=(2\pi m/\beta)^{3N/2}(\beta
P)^{-(N+1)}$ normalizes the distribution. This probability distribution is in
the form of an exponential family with
$(\eta_{1}(\btheta),\eta_{2}(\btheta))=(\beta,\beta P)$,
$(\Phi_{1}(x),\Phi_{2}(x))=(\mathbb{P}^{2}/2m,V)$, $h(x)=1$ and
$A(\btheta)=\log(Z(P,\beta))$. Using the coordinates $(p,\beta)$, where
$p=\beta P$, its FIM is $(ds)^{2}=(N+1)(dp/p)^{2}+(3N/2)(d\beta/\beta)^{2}$. The scalar curvature of the resulting manifold is zero everywhere, implying
that it is a developable surface. 

{\em Known embedding:} By defining a new pair of coordinates  $(x,y)=(\sqrt{1+N}\log(p),\sqrt{3N/2}\log(\beta))$ we have a two dimensional Euclidean embedding. However, the pairwise distance in this embedding is not given by $D^{2}_{sKL}$ and in fact it is not obtainable from any symmetrized R\'enyi divergence~\footnote{The pairwise distance in the 2D Euclidean embedding is $(1+N)\log^{2}{(p_{1}/p_{2})}+(3N/2)\log^{2}{(\beta_{1}/\beta_{2})}$ whereas the symmetrized R\'enyi divergence of ideal gas is $D^{2}_{\alpha}=(1-\alpha)^{-1}(F(\theta_{1})+F(\theta_{2})-F(\alpha\theta_{1}+(1-\alpha)\theta_{2})-F(\alpha\theta_{2}+(1-\alpha)\theta_{2})$, where $\theta=(p,\beta) $ and $F(p,\beta)=(N+1)\log{p}+(3N/2) \log{\beta}$.}. 

IsKL isometrically embeds the ideal gas into (2+2) dimensions. To determine the axis of projection analytically, note that the ideal gas law $PV=N/\beta$ yields the sufficient statistics $\langle \mathbb{P}^{2}/2m\rangle=N/\beta$ and $\langle V\rangle=N/p$. Hence, the pairwise KL divergence between two distributions is
\begin{equation}
\begin{split}
    &D_{sKL}^{2}(p_{1},p_{2},\beta_{1},\beta_{2})\\&=N(p_{1}-p_{2})\bigg(\frac{1}{p_{1}}-\frac{1}{p_{2}}\bigg)+N(\beta_{1}-\beta_{2})\bigg(\frac{1}{\beta_{1}}-\frac{1}{\beta_{2}}\bigg).
\end{split}
\end{equation}
Letting the centers of mass be $\overline{\langle \eta\rangle}=\langle \eta\rangle$ and $\overline{\langle \Phi\rangle}=\langle \Phi\rangle$, the projection coordinates are given by 
\begin{equation}
    \begin{split}
        T_{p}^{\pm}(p)&=\frac{1}{2}\bigg(\lambda_{p} \Big(\,p-\big\langle p\big\rangle\Big) \pm N\lambda_{p}^{-1}\Big(\,p^{-1}-\big\langle p^{-1} \big\rangle\Big)\bigg)\\
        T_{\beta}^{\pm}(\beta)&=\frac{1}{2}\bigg(\lambda_{\beta} \Big(\,\beta-\big\langle \beta\big\rangle\Big) \pm N\lambda_{\beta}^{-1}\Big(\,\beta^{-1}-\big\langle \beta^{-1} \big\rangle\Big)\bigg)
        \label{eq:coord_ideal}
    \end{split}
\end{equation} 
From Eq. \ref{eq:coord_ideal}, the coordinate pairs yield
$(T_{k}^{+}-C^{+}_{k})^{2}-(T_{k}^{-}-C^{-}_{k})^{2}=r^{2}$, where
$k=\{p,\beta\}$, $r^{2}=N$ and $C^{\pm}_{k}=(1/2)\big(-\lambda_{k}\langle
k\rangle\pm N\lambda^{-1}_{k}\langle
k^{-1}\rangle\big)$ are constants that depend on the sampling
range. Therefore, the ideal gas manifold is a four dimensional Minkowskian
torus (topologically a hyperboloid) with radii $r_{1} =r_{2}=\sqrt{N}$. Its projections are illustrated in Fig.~\ref{fig:gas}(b). Just as
the 4D Euclidean torus has zero curvature \footnote{The Gauss Bonnet theorem
tells us that the integral of the torus curvature is zero, the 4D torus
$\mathbb{S}^{1}\times\mathbb{S}^{1}$ has a zero curvature.}, so it does in
Minkowski space. 

We can map our \isKL\ embedding onto the known embedding into $\mathbb{R}^2$
above. Roughly speaking, this works because our torus is 
the Cartesian product of two circles with zero Gaussian curvature.
We are thus able to provide a mapping to the Euclidean embedding discussed by shifting the coordinates, $T_{k}^{\pm}\rightarrow T_{k}^{\pm}-C^{\pm}_{k}$ and  parameterizing the coordinate pairs as $(T_{k}^{+},T_{k}^{-})=(\sqrt{N}\cosh(\phi_{k}),\sqrt{N}\sinh(\phi_{k}))$, where  $\phi_{k}=(1/2)\log(k\lambda_{k}/\sqrt{N})$ and $k\in\{p,\beta\}$. This gives  
\begin{equation}
\begin{split}
    (x,y)&=\bigg(\sqrt{1+N}\bigg(\log\bigg(\frac{\sqrt{N}}{\lambda_{p}}\bigg)+2\phi_{p}\bigg),\\&\sqrt{\frac{3N}{2}}\bigg(\log\bigg(\frac{\sqrt{N}}{\lambda_{\beta}}\bigg)+2\phi_{\beta}\bigg)
\end{split}
\end{equation}
where the 'circles' have been unwound to straight lines through the hyperbolic angle $\phi_{k}$. 

 Fig.~\ref{fig:gas}(a) shows the three dimensional projection of the ideal gas manifold which is colored based on the inverse temperature $\beta$. Discussion of the ideal gas is
often accompanied by that of the thermodynamic cycles with which it can be used
to extract work from a heat bath. The Carnot cycle, which is often considered
to cost no entropy, was recently shown~\cite{machta2015dissipation} to have a
sub-extensive entropy cost proportional to the
arc length of the cycle's path on the model manifold. This challenges Szilard's
argument that information entropy and thermodynamic entropy can be freely
exchanged. The path of a Carnot cycle is shown on the model manifold in
Fig. \ref{fig:gas}(a).
\subsection{The n-sided die}
\label{sec:dice}
\begin{figure}
    \centering
    \includegraphics[scale=0.46]{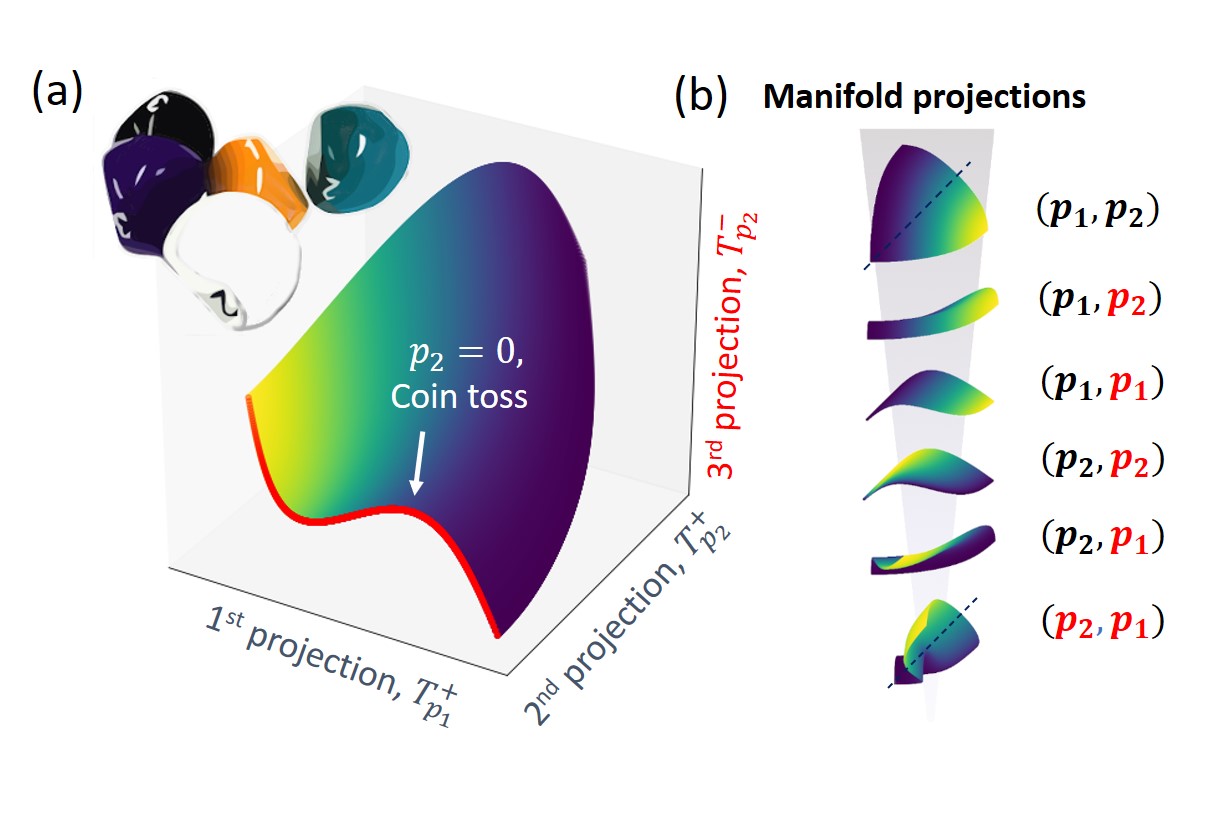}
    \caption{{\bf Model manifold for the three sided die} is embedded into (2+2) dimension with isKL embedding. a) The three dimensional projection of the three sided die manifold is colored according to the fairness parameter $p_{1}$. Depicted also is the coin toss submanifold in red. b) The manifold projections are arranged based on the manifold widths. The spacelike directions are color-coded in black while the timelike directions are color-coded in red. The analytical expression for each projection is given by Eq.~\eqref{eq:dice3}. We have permutation symmetry in $(T^{+}_{p_{i}},T^{-}_{p_{j}})$  coordinate pairs and reflection symmetry along the $p_{1}=p_{2}$ line (dotted line) in ($T^{\pm}_{p_{i}},T^{\pm}_{p_{j}})$ coordinate pairs.}
    \label{fig:dice}
\end{figure}

The $n$ sided die is a model for a process with $n$ outcomes. It has a discrete
probability distribution of $n$ states, with $p_{i}$ as the probability of the
$i$th state. This distribution can be written as
$P(x|\boldsymbol{p})=\prod_{i=1}^{n}p_{i}^{[x=i]}$, where $[x=i]$ is the Iverson
bracket which evaluates to 1 if $x=i$, 0 otherwise and $\sum_{i=1}^{n}p_{i}=1$. The probability distribution
can be written in the form of an exponential family with
$\eta_{i}(\btheta)=\log(p_{i}/p_{n})$, $\Phi_{i}=[x]$, $h(x)=1$ and
$A(\boldsymbol{\theta})=\log(1+\sum_{i=1}^{n-1}e^{\theta_{i}})$. Its FIM is
$(ds)^{2}=\sum_{i=1}^{n}(dp_{i})^{2}/p_{i}$.

{\em Known embedding:} Taking $\sqrt{p_{i}}$ as parameters instead of $p_i$ gives an embedding onto a
Hellinger $n$-sphere. This implies that in the Hellinger embedding, the $n$
sided die manifold has both permutation and spherical symmetry. Moreover, since
this mapping is a universal cover of $n$-sphere its scalar curvature must be
positive \cite{wolf1962homogeneous}. For example, the scalar curvature of a
three sided die and a four sided die are 1/2 and 2 respectively. 

IsKL produces an embedding in $(n-1)+(n-1)$ dimensions. As $\langle \Phi_{i}\rangle=p_{i}$, the pairwise KL divergence between $P_{\boldsymbol{p}}$ and $P_{\boldsymbol{a}}$ is
\begin{equation}
D^{2}_{sKL}(\textbf{p},\textbf{a})=\sum_{i=1}^{n}(p_{i}-a_{i})\log\bigg(\frac{p_{i}}{a_{i}}\bigg).
\end{equation}
By letting $\overline{\langle \eta_{i}\rangle}=\langle\eta_{i} \rangle$ and $\overline{\langle \Phi_{i}\rangle}=\langle\Phi_{i} \rangle$, the projection coordinates are
\begin{equation}
\begin{split}
    &T_{k}^{\pm}(p_{1},...,p_{n-1})\\&=\frac{1}{2}\bigg(\lambda_{k}\bigg(\log{\bigg(\frac{p_{k}}{p_{n}}\bigg)}-\bigg\langle{\log{\bigg(\frac{p_{k}}{p_{n}}\bigg)}\bigg\rangle}\bigg)\pm\frac{1}{\lambda_{k}}\big(p_{k}-\langle{p}\rangle\big)\bigg)
\end{split}
\end{equation}
where $k=1,...,n-1$ and $p_{n}=1-\sum_{i=1}^{n-1}p_{i}$.
As examples, we consider three and four sided dice.  IsKL gives (2+2) and (3+3) dimensional embeddings in Minkowski space. There are only two eigenvalues returned in both cases, signalling the existence of symmetries in our embeddings. With uniform sampling of the parameter space, for $n=3$, 
\begin{equation}
\begin{split}
T^{(k)}_{\pm}(p_{1},p_{2})&=\frac{1}{6^{1/4}\sqrt{\pi}} \log{\bigg(\frac{p_{k}}{p_{3}}\bigg)}\pm6^{1/4}\sqrt{\pi}\bigg(p_{k}-\frac{1}{3}\bigg)
\end{split}
\label{eq:dice3}
\end{equation}
where $k=1,2$ and $p_{3}=1-p_{1}-p_{2}$. For $n=4$, 
\begin{equation}
\begin{split}
&T^{(k)}_{\pm}(p_{1},p_{2},p_{3})\\&=\frac{1}{5^{1/4}}\sqrt{\frac{3}{4\pi}} \log{\bigg(\frac{p_{k}}{p_{4}}\bigg)}\pm5^{1/4}\sqrt{\frac{4\pi}{3}}\bigg(p_{k}-\frac{1}{4}\bigg)
\end{split}
\end{equation}
where $k=1,2,3$ and $p_{4}=1-p_{1}-p_{2}-p_{3}$. Finally, the projection coordinates for $n=2$ (a coin toss) are
\begin{equation}
\begin{split}
T^{(2)}_{\pm}(p)&=\frac{1}{\sqrt{2\pi}}\log\bigg(\frac{p}{1-p}\bigg)\pm\sqrt{2\pi}\bigg(p-\frac{1}{2}\bigg).\\
\end{split}
\label{eq:coin_uni}
\end{equation}
As expected, comparing Eq. \eqref{eq:coin_uni} with Eq. \eqref{eq:coinflip_skl}, the form does not depend on the sampling choice while the constant $\lambda_{p}$ does.
 Fig.~\ref{fig:dice}(b) shows the numerically calculated manifold projections. The manifold is coloured based on the fairness parameter $p_{1}$.  Unlike the
 Hellinger embedding, the lack of spherical symmetry is manifest. We do however
 see a permutation symmetry among $p_{i}$s and a reflection symmetry along
 $T^{\pm}_{p_{1}}=T^{\pm}_{p_{2}}$  in the $(T^{\pm}_{p_{1}},T^{\pm}_{p_{2}})$
 coordinate pairs. One can extract the sub-manifold of a coin toss problem by
 restricting $p_{2}=0$. This submanifold is shown by the red line in
 Fig.~\ref{fig:dice}(a). In general, any discrete probability distribution is a
 subset of the $n$ sided die distribution, implying that other discrete
 exponential family distributions may have hidden low dimensional
 representation within the $n$ sided die model manifold. 
  
\subsection{Nonlinear least square models}
\label{subsec:NonlinearLeastSquares}

Nonlinear least square models are ubiquitous in fitting deterministic models to
data with noise. These models take the form of a nonlinear vector-valued
function $f_{i}(\btheta)$ predicting the value of experimental data points
$x_{i}$ with uncertainties $\sigma_{i}$. Their associated probability distribution is
\begin{equation}
P(\boldsymbol{x}|\btheta) = \prod_i \frac{1}{\sqrt{2\pi\sigma_i^2}}
		\exp\left(\frac{(f_i(\btheta)-x_i)^2}{2\sigma_i^2}\right).
\end{equation}
This probability distribution takes the form of an exponential family
with $\eta_{i}(\btheta)=f_{i}(\btheta)/\sigma_{i}$,
$\Phi(x_{i})=x_{i}/\sigma_{i}$, 
$h(\boldsymbol{x})= \sum_{i}x^{2}_{i}/\sigma^{2}_{i}$ and 
$A(\btheta)=\sum_{i} f_{i}^{2}(\btheta)/2\sigma_i^{2}-\log(2\pi \sigma_i^2)/2 $.
Unlike the other models discussed, which have the same number of natural parameters $\eta_i(\btheta)$ and model parameters
$\theta_i$, here the number of natural parameters is given by the number of data points being fit. The FIM is given by 
$J^{\top}_{\beta i} J_{i\alpha}$, where
$J_{i\alpha}=\partial f_{i}(\btheta)/\partial \theta_{\alpha}$ is the Jacobian.

\begin{figure}
    \centering
    \includegraphics[scale=0.33]{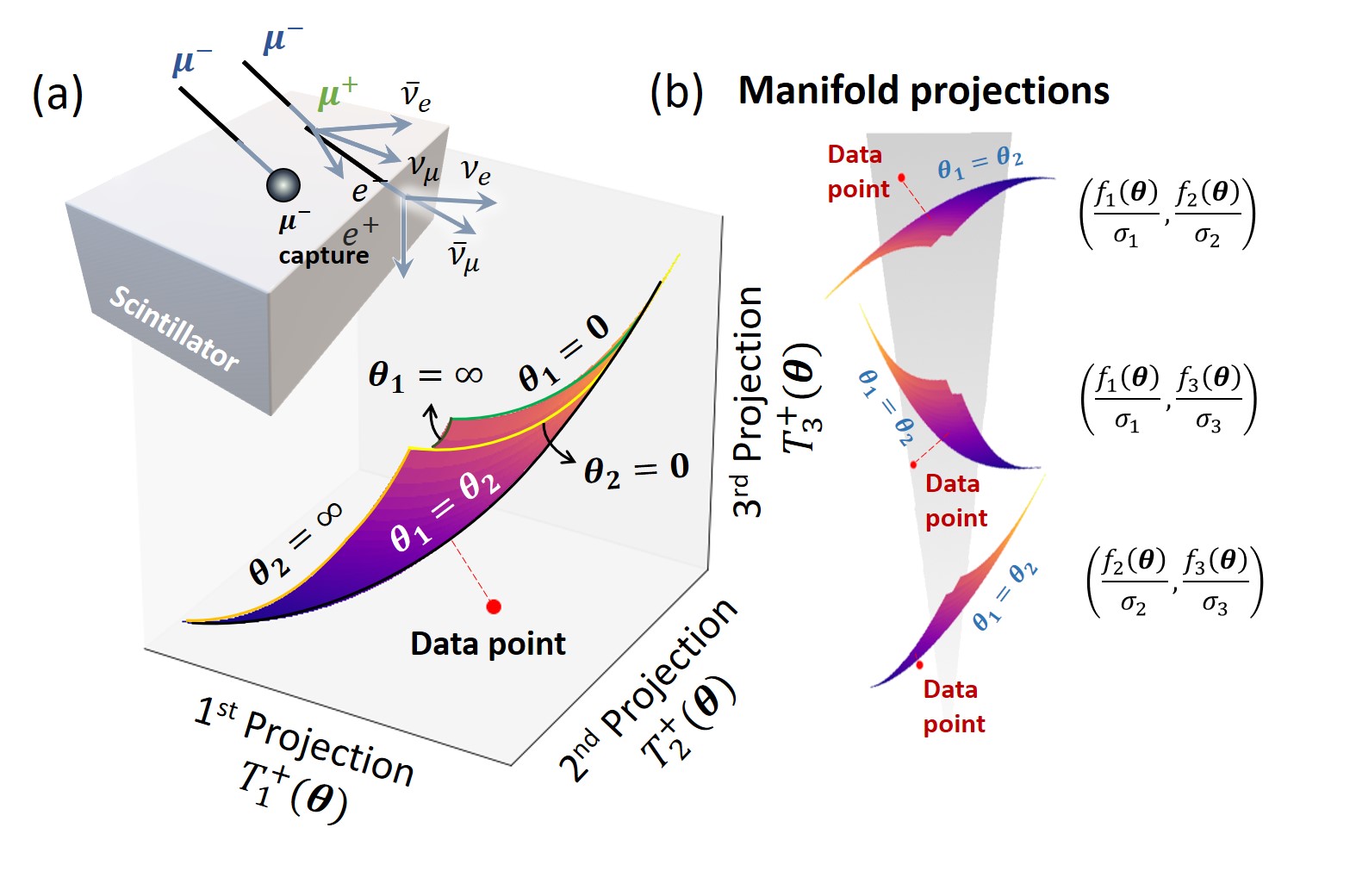}
    \caption{\textbf{Model manifold for the muon lifetime}, our two-parameter
least-squares model, evaluated at three time points. The \isKL\ embedding
is confined to three Euclidean dimensions, with the three time-like coordinates
identically zero. a) The manifold is colored with the muon lifetime $\theta$. The model manifold is bounded with four edges at
$\theta_{k}=0$ and $\theta_{k}=\infty$ and a tight fold
along $\theta_1 = \theta_2$. 
Depicted also is the experimental data point in red which is in close 
proximity to the $\theta_{1}=\theta_{2}$ boundary See~\cite[Fig.~1]{TranstrumMS10}. b) The manifold projections of the muon lifetime model manifold are arranged based on the manifold widths. The analytical expression for each axis is given by Eq.~\ref{eq:muon_eq}.
}
    \label{fig:nls}
\end{figure}

{\em Known embedding:} Least-squares models with $N$ data points have a natural 
`prediction embedding' into $N$-dimensional
Euclidean space with one coordinate per data point $x_i$ given
by the error-normalized model prediction $f_i(\btheta)/\sigma_i$. 
While the number of data points can be much larger than the number of 
parameters, this embedding remains valuable because the model predictions
are surprisingly often well approximated by low-dimensional, flat model
manifolds we call 
{\em hyperribbons}~\cite{TranstrumMS10,TranstrumMS11,QuinnWTS19}.
Hyperribbons have a hierarchy of manifold widths---like a ribbon, their dimensions
(length, width, thickness, \dots) become geometrically smaller---yielding
predictions that depend mostly on the first few principal components. 
Our least-squares model has $N$ natural parameters, so isKL will produce
an embedding into an $N+N$ dimensional Minkowski space. Can we find one
that makes the time-like distances equal to zero, reproducing the 
$N$-dimensional prediction embedding?

The symmetrized Kullback-Leibler divergence between two models is
indeed given by the Euclidean distance between the two model predictions:
\begin{equation}
    D^{2}_{sKL}(\btheta_{1},\btheta_{2})=\sum_{i=1}^{N}\frac{(f_{i}(\btheta_{1})-f_{i}(\btheta_{2}))^{2}}{\sigma_{i}^{2}}.
    \label{eq:nls}
\end{equation}
This appears promising: the isKL distance is the same as that of the 
prediction embedding above. Interestingly, any \Renyi\ divergence (such
as the Bhattacharyya distance used by \inPCA~\cite{QuinnCdBNS19}) gives
the same pairwise distance measure.
Since $\langle \Phi(x_{i}) \rangle =f_{i}(\btheta)/\sigma$, the
projection coordinates are
\begin{equation}
T_{i}^{\pm}(\btheta)=\frac{1}{2\sigma_{i}}\bigg(\lambda\pm\frac{1}{\lambda}\bigg)
\bigg(f_{i}(\btheta)-\langle f_{i}(\btheta)\rangle\bigg)
\label{eq:muon_eq}
\end{equation}
By taking $\lambda=1$ the time-like coordinates vanish and we reproduce
the $N$-dimensional prediction embedding.

Figure~\ref{fig:nls} shows this prediction embedding for the classical
nonlinear least squares model of two exponential decays, here in the context of a 
cosmic muon lifetime experiment. Approximately half of muons generated by cosmic ray collisions are
negative muons which can be captured by a proton of host nuclei.
The effective negative muon lifetime $1/\theta_2$ (including capture) is therefore
expected to be shorter than the decay-only lifetime of positive
muons $1/\theta_1$. The model prediction for the number of muons surviving after
some time $N(t)$ is thus the sum of two exponentials. Mathematically, we have
\begin{equation}
\hat{N}(\theta_{1},\theta_{2},r ,t)
  =\frac{1}{1+r}\big(r e^{-\theta_{1} t}+e^{-\theta_{2}t}\big)
\label{eq:decay}
\end{equation}
where $\hat{N}(t)$ is the normalized number of muons and $r = N_{\mu^{+}}/N_{\mu^{-}} = 1.18 \pm 0.12$ is the
ratio of incident positive muons to negative muons formed by the cosmic
rays \cite{morewitz1953variation}. Fig.~\ref{fig:nls} shows the muon lifetime model manifold via the
isKL embedding (identical to the prediction embedding), with three
sampled time points. The manifold is colored based on the muon lifetime $\theta_{1}$. The projection coordinates are
$\hat{N}(t_{i})/\sigma_{i}$. 
%Visually, the model manifold has boundaries
%formed by the limiting cases 
%i)~$\theta_{i}=0$, 
%ii)~$\theta_{1}=\theta_{2}$ and \hb
%iii)~$\theta_{i}=\infty$.
%\han{yes there are two layers...}{\bf XXX Something is wrong here! %If $r !=1$, there is no symmetry when
%$\theta_1$ swaps with $\theta_2$. I'm guessing there are two layers %to the
%model manifold in the figure!}\hb
Since $r\approx 1$, there is a tight fold in the model manifold along
$\theta_{1}=\theta_{2}$. The experimental data point is close to the
manifold fold, implying the negative muon capture event only leads 
to a slight change in negative muon lifetime.

\subsection{Gaussian fits to data}
\label{sec:gaussfit}
\begin{figure}
    \centering
    \includegraphics[scale=0.38]{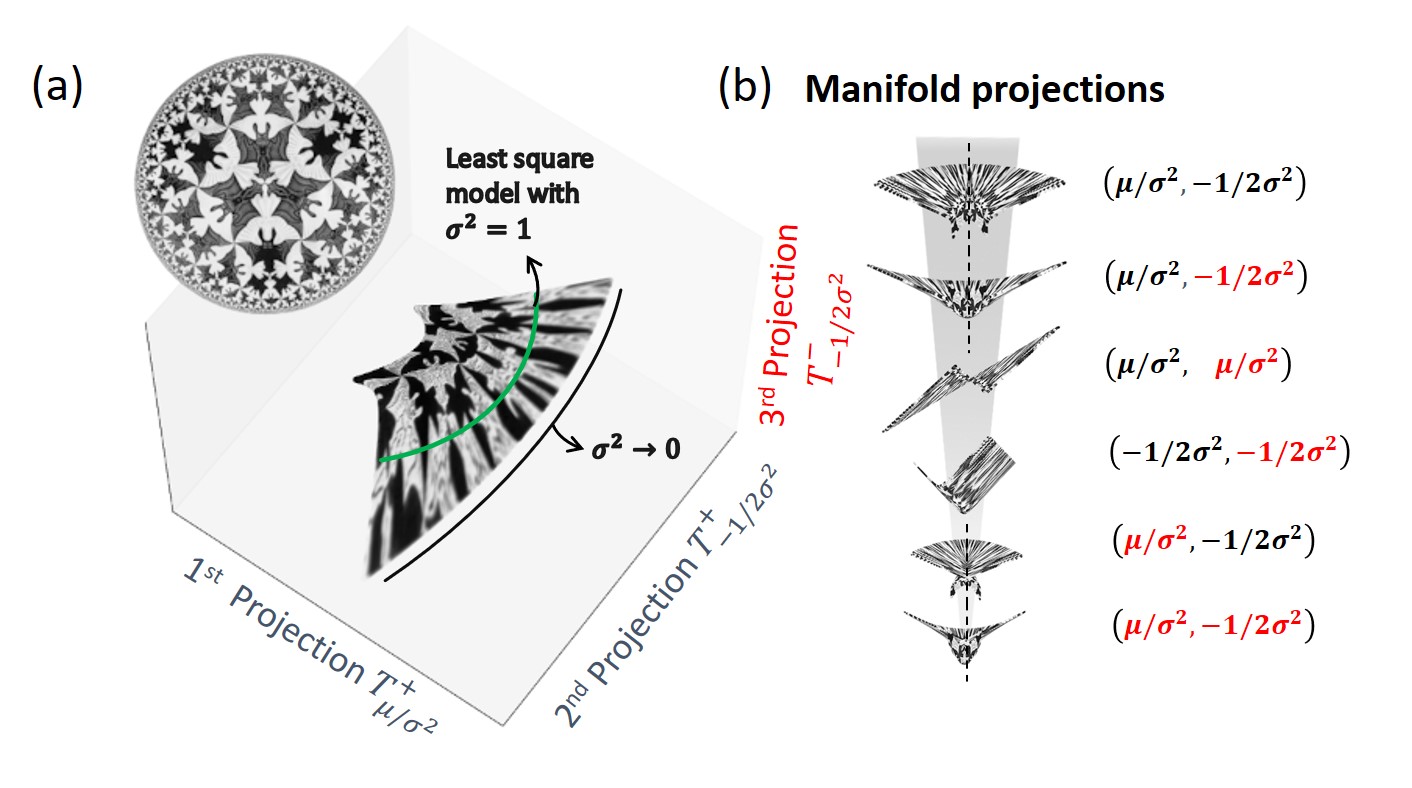}
    \caption{{\bf Viewing Heaven and Hell in Minkowski Space ---} The Gaussian fit manifold is embedded into (2+2) dimension with isKL embedding. a) The three dimensional projection of the Gaussian fit manifold is decorated with Escher's art ---Circle Limit IV which is also known as  Heaven and Hell. The submanifold of a least square model with a single Gaussian distribution of fixed $\sigma^{2}=1$ is depicted in green. b) The manifold projections are depicted in a descending order based on the manifold widths along the spacelike/timelike components. The spacelike directions are color-coded in black while the timelike directions are color-coded in red. The analytical expression for each axis is given by Eq.~\eqref{eq:coord_gauss}.  Reflection symmetry is illustrated with a dashed line along projections with a $\mu/\sigma^{2}$ component.   }
    \label{fig:gauss}
\end{figure}
. \begin{figure*}
    \centering
    \includegraphics[scale=0.45]{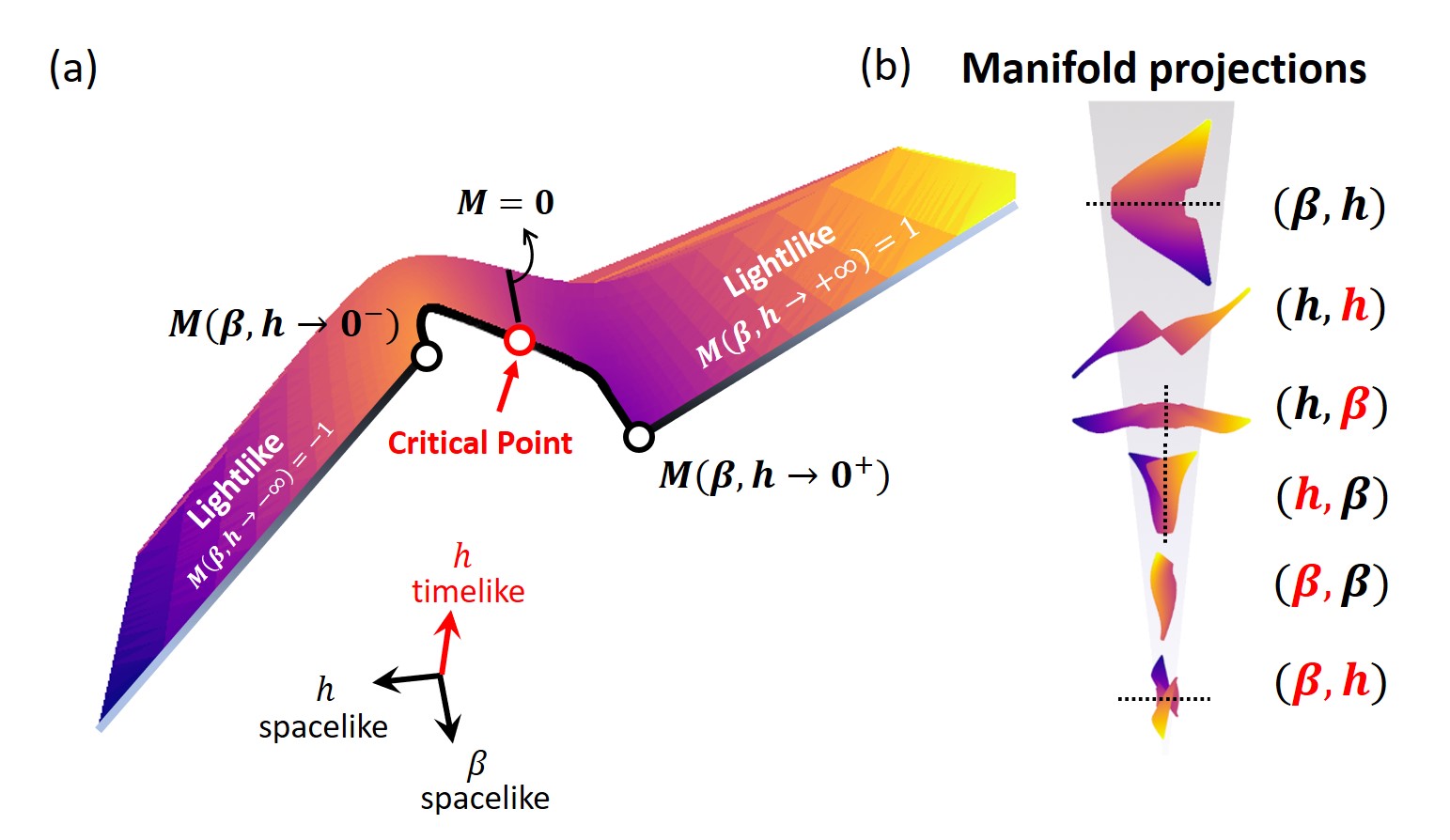}
    \caption{
      {\bf Two dimensional Ising Model isKL embedding} is used to illustrate
      the geometric structure of statistical models with a phase transition. The Ising model manifold is embedded into (2+2) dimensions.  a) The three dimensional projection of the Ising model manifold is colored based on the external magnetic field $h$. For $\beta>\beta_{c}$, there is
      an opening on the manifold due to the spontaneous magnetization. The two arms illustrated
      correspond to magnetization $M(\beta,h)=\pm 1$ with $\beta>\beta_{c}$ are
      lightlike. The values of Ising average energy $E$ and magnetization  $M$ used were estimated from simulations
      with $n=128\times128$ spins. The exact solution at zero field is depicted by the black line. b)
      The Ising model manifold
      projections are shown in a descending order based on the manifold widths
      along the spacelike/timelike directions. The spacelike direction are
      color-coded in black while the timelike directions are color-coded in
      red. The analytical expression for each axis of projection is given by Eq.~\eqref{eq:ising}. Reflection symmetry is illustrated with a dotted line along
      projections with a external magnetic field $h$ component.  
    }
   \label{fig:ising2d}
\end{figure*}

The Gaussian distribution is an exceptionally good approximation for many physical problems and thus serves as a good model to explore in the context of manifold visualization. For example the distribution of women's heights with mean height $\mu$ and variance in height $\sigma^{2}$ in a country is fitted to a normal (Gaussian) distribution.
The Gaussian distribution $P(x|\mu,\sigma)=(2\pi\sigma^{2})^{-1/2}\exp(-(x-\mu)^{2}/2\sigma^{2})$ has two parameters, the mean $\mu$ and the variance $\sigma^{2}$.  It can be written in the form of an exponential family with $(\eta_{1}(\btheta),\eta_{2}(\btheta))=(\mu/\sigma^{2},-1/2\sigma^{2})$, $(\Phi_{1}(x),\Phi_{2}(x))=(x,x^{2})$, $h(x)=(2\pi)^{-1/2}$ and $A(\theta_{1},\theta_{2})=-\theta_{1}^{2}/4\theta_{2}-(1/2)\log{(-2\theta_{2})}$. Its FIM is given by $(ds)^{2}=\sigma^{-2}((d\mu)^{2}+2(d\sigma)^{2})$.

{\em Known embeddings:} The Gaussian distribution FIM has a close resemblance to the Poincare half plane metric $(ds)^{2}=y^{-2}((dx)^{2}+(dy)^{2})$ both of which have a constant negative scalar curvature: -1/2 and -2, respectively.  
In differential geometry, it is known~\cite{guan2007isometric} that the Poincar\'e half plane has an isometric canonical embedding into (2+1) dimensional Minkowski space and takes the form of an imaginary sphere with radius squared equal to minus one. By rescaling, the corresponding embedding for the Gaussian fit manifold is therefore an imaginary sphere of radius squared equal to -2. Its spacelike components are given by $ X^{+}_{1}(\mu,\sigma)=(\mu^{2}+2\sigma^{2}+2)/2\sqrt{2}\sigma^{2}$, $X^{+}_{2}(\mu,\sigma)=\mu/\sigma$ and its timelike component is given by  $X^{-}_{3}(\mu,\sigma)=(\mu^{2}+2\sigma^{2}-2)/(2\sqrt{2}\sigma^{2})$.
The pairwise distance which generates such an embedding is therefore
\begin{equation}
    D^{2}(\mu_{1},\sigma_{1},\mu_{2},\sigma_{2})=\frac{(\mu_{1}-\mu_{2})^{2}+2(\sigma_{1}-\sigma_{2})^{2}}{2\sigma_{1}\sigma_{2}}
    \label{eq:gausspd}
\end{equation}
However, there is no obvious way of writing Eq. \eqref{eq:gausspd} in terms of $P(x|\mu,\sigma)$. 

With the isKL embedding, the Gaussian distribution can be isometrically embedded into (2+2) dimensions. As $\langle \Phi_{1}(x)\rangle=\mu$ and $\langle \Phi_{2}(x)\rangle=\mu^{2}+\sigma^{2}$, the pairwise distance is given by
\begin{equation}
\begin{split}
D^{2}_{sKL}(\mu_{1},\mu_{2},\sigma^{2}_{1},\sigma^{2}_{2})&=\bigg(\frac{\mu_{1}}{\sigma^{2}_{1}}-\frac{\mu_{2}}{\sigma_{2}^{2}}\bigg)(\mu_{1}-\mu_{2})\\-&\frac{1}{2}\bigg(\frac{1}{\sigma^{2}_{1}}-\frac{1}{\sigma^{2}_{2}}\bigg)\big(\mu^{2}_{1}+\sigma^{2}_{1}-\mu^{2}_{2}-\sigma^{2}_{2}\big)
\end{split}
\end{equation}
Letting $\overline{\langle\eta\rangle}=\langle \eta\rangle$ and $\overline{\langle\Phi\rangle}=\big\langle \Phi\big\rangle$, the coordinates are given by
\begin{widetext}
\begin{equation}
\begin{split}
    T^{\pm}_{odd}(\mu,\sigma^{2})&=\frac{1}{2}\left(\lambda_{odd}\bigg(\,\frac{\mu}{\sigma^{2}}-\bigg\langle \frac{\mu}{\sigma^{2}} \bigg\rangle\bigg)\pm\frac{1}{\lambda_{odd}}\big(\,\mu-\langle \mu \rangle\big)\right)\\
    T^{\pm}_{even}(\mu,\sigma^{2})&=\frac{1}{2}\bigg(\lambda_{even}\bigg(\,\frac{1}{\sigma^{2}}-\bigg\langle \frac{1}{\sigma^{2}} \bigg\rangle\bigg) \pm\frac{1}{\lambda_{even}}\big(\,\mu^{2}+\sigma^{2}-\langle \mu^{2}+\sigma^{2} \rangle\big)\bigg).
    \label{eq:coord_gauss}
\end{split}
\end{equation}
\end{widetext}

Upon closer inspection, the coordinate pairs can be written as
\begin{equation}
\begin{split}
&(T_{odd}^{+}-C^{+}_{odd})^{2}-(T_{odd}^{-}-C^{-}_{odd})^{2}\\&-(T_{even}^{+}-C^{+}_{even})^{2}+(T_{even}^{-}-C^{-}_{even})^{2}=1 
\end{split}
\end{equation} 
where $C^{\pm}$ are constants. This suggests the isKL embedding is a 4 dimensional hyperboloid in Minkowski space. 
 To get a good pictorial sense of how the probability distributions are
 arranged, we embedded 'Heaven and Hell' (Escher's Circle Limit IV 1960-
 depicting a Poincare disk) in Minkowski space via our isKL embedding. Fig.~\ref{fig:gauss}(a) shows the three dimensional projection of the manifold and in Fig.~\ref{fig:gauss}(b), the manifold projections along the spacelike (black) and timelike (red) axes are to scale and accurately capture the manifold widths. The probabilistic manifold projection along
 $(\mu/\sigma^{2},-1/2\sigma^{2}),(\mu/\sigma^{2},\textcolor{red}{-1/2\sigma^{2}}),
 (-1/2\sigma^{2},\textcolor{red}{\mu/\sigma^{2}})$ and
 $(\textcolor{red}{-1/2\sigma^{2}},\textcolor{red}{\mu/\sigma^{2}})$ components
 exhibit a reflection symmetry about $\mu=0$, manifesting the even parity
 coordinates. Moreover, the bats become stretched as $\sigma^{2}\rightarrow0$,
 along the projected edge of the Poincar\'e disk. The
 submanifold of a least square model with a single Gaussian distribution of
 fixed $\sigma^{2}=1$ from Sec. \ref{sec:curseofdim} in shown in green
\subsection{2D Ising model }
\label{sec:ising}
\begin{figure*}
    \centering
    \includegraphics[scale=0.45]{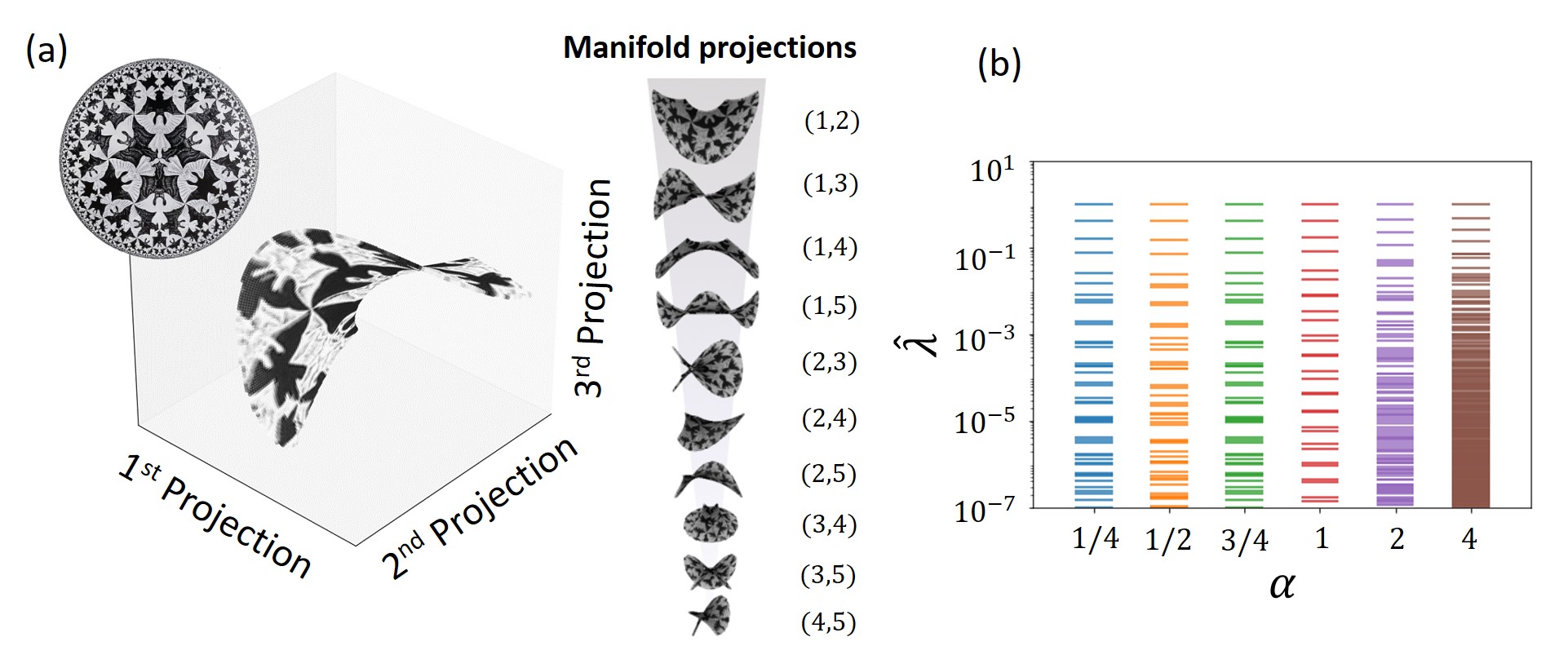}
    \caption{{\bf The Cauchy distribution} is considered to exemplify the rough equivalence of the isKL embedding with various other Minkowski embeddings for visualizing non exponential family distributions. a) The three dimensional projection of the Cauchy distribution manifold is shown on the left. To compare it with the Gaussian fits manifold, we have colored the Cauchy manifold with Escher's art - Circle Limit IV. Here, the bat shapes are well preserved as compared to Fig.~\ref{fig:gauss}. The first 5 manifold projections are shown on the right in a descending order based on the manifold widths along the $(m,n)$ principal components. b) Squared principal length of intensive embedding with different symmetrized R\'enyi divergences for the Cauchy manifold. Here, we observe a geometrically decreasing manifold widths that spans many decades for all $\alpha$s.  } 
    \label{fig:cauchy}
\end{figure*}
Most statistical mechanics models form exponential families, and of
particular interest is the behavior of their model manifolds near phase
transitions. Here we show how the two dimensional Ising model manifold is
embedded using our method. The Ising model is a model of magnetism comprised of
a lattice of $n$ spins that can take the values $\pm1$, ``pointing up'' or
``pointing down.'' At temperature $\beta^{-1}$ and in an external magnetic
field $H$, the probability of observing a particular configuration $\boldsymbol
s=(s_1,\ldots,s_n)$ of the spins is given by the Boltzmann distribution
\begin{equation}
    P(\boldsymbol{s}|\beta, h)=\frac{\exp{(\beta\sum_{\langle ij\rangle}s_{i}s_{j}+h\sum_{i} s_{i}})}{Z(\beta,h)}
\end{equation}
where $h=\beta H$, $\langle ij\rangle$ denotes a sum over neighbouring sites,
and the partition function $Z(\beta, h)$ normalizes the distribution. The Ising model is an exponential family with 
$(\eta_{1}(\btheta),\eta_{2}(\btheta))=(\beta, h)$,
$(\Phi_{1}(\boldsymbol{s}),\Phi_{2}(\boldsymbol{s}))=(\sum_{\langle
ij\rangle}s_{i}s_{j},\sum_{i}s_{i})$, $h(\boldsymbol{s})=1$, and
$A(\boldsymbol{\theta})=-\log Z$.  The Fisher information metric is given by
the mixed partial derivatives $g_{ij}=\partial_{i}\partial_{j}\log Z$ with
$i,j\in\{\beta,h\}$. 

{\em Known embeddings:} The Hellinger embedding of the Ising model manifold is $2^{n}$ dimensional. The
curse of dimensionality manifests through an increase of `wrapping' around the
unit hypersphere as the number of spins increases, rendering low dimensional
projections increasingly useless for visualization~\cite{QuinnWTS19}. The
`wrapping' phenomenon can be ameliorated by using the InPCA embedding. Though InPCA still
embeds the Ising model manifold in a high dimensional Minkowski space,
the length scales of adjacent principal components are well-separated. 

IsKL embeds
the Ising model manifold into (2+2) dimensions. Not only is the curse of
dimensionality broken, the Ising model manifold is embedded into \emph{finite}
dimensional Minkowski space. The expectation values of the sufficient
statistics can be related directly to the Ising average energy $E$ and
magnetization $M$ by
$(\langle\Phi_{1}\rangle,\langle\Phi_{2}\rangle)=(HM-E,M)$. The pairwise
distance is then % \jaron{You previously had $\langle\Phi_1\rangle$ with the wrong sign, which switches $T_\beta^+$ and $T_\beta^-$. The plots should be adjusted accordingly...}
\begin{widetext}
\begin{equation}
    D^{2}_{sKL}(\beta_{1},\beta_{2},h_{1}, h_{2})= (\beta_{2}-\beta_{1})(M_{2}h_{2}/\beta_2-E_2-M_{1}h_{1}/\beta_1+E_1) +(h_{2}-h_{1})(M_{2}-M_{1})
\end{equation}
\end{widetext}
The Ising model manifold is centered at the critical point
$(\beta,h)=(\beta_{c},0)$ with the projection coordinates being
\begin{equation}
\begin{split}
    T^{\pm}_{\beta}&=\frac{1}{2}\bigg(\lambda_{\beta} (\beta-\beta_{c})\pm\frac{1}{\lambda_{\beta}}(Mh/\beta-E+E_c)\bigg)\\
    T^{\pm}_{h}&=\frac{1}{2}\bigg(\lambda_{h} h\pm\frac{1}{\lambda_{h}}M\bigg)
\label{eq:ising}
\end{split}
\end{equation}
where $E_c$ is the average energy at the critical point.
Fig.~\ref{fig:ising2d} shows the isKL embedding of the 2D Ising manifold with
$E$ and $M$ estimated from Monte Carlo simulations at $n=128\times128$ spins
% Rejection-free is important here
using a rejection-free variant of the Wolff algorithm in an external
field~\cite{kent2018cluster}. The
exact solution for the zero field is included in the embedding as well and is
illustrated with a black line \cite{onsager1944crystal, yang1952spontaneous}
%\han{Added}\jaron{Can we got a black line on the plots for the %\emph{high temperature}
%exact solution?}. 
For completeness we
also show all the manifold projections. The first and third principal
components are field like directions and the 2nd and the 4th components are
temperature like directions. Reflection symmetry along $H=0$ is depicted with a
dotted line. This observation is further highlighted by having the Ising model manifold colored based on the external magnetic field $h$. 

At the critical point there is an opening that corresponds to the growing
spontaneous magnetization. This resolves a serious-seeming problem with any embedding based on the Fisher information metric. The FIM can be written in terms of the free energy, and the free energies for the two zero-field branches $\pm M(T)$ agree: the two magnetizations are zero distance apart, even though they manifestly are far apart in probability space. Any Euclidean embedding will place them at the same point. The embedding in Minkowski space resolves this: the zero distributional distance manifests itself in a large, physically sensible opening in the embedding, along a line of light-like separation. This highlights
the crucial role of timelike coordinates in qualitatively differentiating
unlike systems that have the same free energy. This is not the whole story of
lightlike separations, however: the two arms highlighted at large $\beta$ in
Fig. \ref{fig:ising2d} are also light-like. These have a more conventional
interpretation: for sufficiently high field the configuration with all
spins in the direction of the field becomes the most probable, and the
resulting distributions are difficult to distinguish. IsKL spreads these points out as well. 

The connection between phase transitions and differential geometry has been
widely
investigated~\cite{brody1995geometrical,janyszek1990riemannian,ruppeiner1979thermodynamics,ruppeiner1995riemannian}.
Researchers have argued that the scalar curvature $R$ can be viewed as a measurement of
interactions and that the divergence of the scalar curvature signals a phase transition. The
leading singularity in the scalar curvature of the 2D Ising model manifold as the
critical point is approached can be computed from the metric above and the asymptotic scaling form $-\log Z\simeq t^2\mathcal F(ht^{-15/8})+t^2\log t^2$ for $t=\beta_c-\beta$ to be
$R\sim-t^{-2}/\log(t^2)$. For small $\beta-\beta_c$
$R$ diverges. Near the critical point one might
expect to see a cusp as a result. Instead, there is an opening near the
critical point in our embedding, and the surrounding manifold looks smooth. The identification of each point along the opening with an opposing point suggests that we may have disguised the cusp in our embedding by `cutting' the manifold with lightlike displacements, the way one might remove the point of a cone by cutting up the side. The connection between the geometry of our
manifold and the singularity of its scalar curvature will be further explored in future
work.

\section{Non-Exponential Families : Cauchy Distribution}

The success of the isKL embedding in obtaining an analytical expression for each coordinate is special to exponential family distributions. As an example of a non-exponential family, we consider the long tailed Cauchy distribution,
\begin{equation}
    P(x|x_{0},\gamma)=\frac{\gamma}{\pi(\gamma^{2}+(x-x_{0})^{2})}.
\end{equation}
Interestingly, its FIM,  $(ds)^{2}=(2\gamma^{2})^{-1}((dx_{0})^{2}+(d\gamma)^{2})$ has a constant negative scalar curvature just as the Gaussian fit in Sec. IV (b).
% In fact, there is a connection between the Gaussian and Cauchy distributions : they belong to the location scale family $p(x|\theta)=\sigma ^{-1}p((x-\mu/)\sigma)$, where $\mu$ the location parameter and $\sigma$ is the scaling factor. It is a known fact any location scale distribution can be embedded to the Poincare half plane[cite XXX]. With the definition of location scale family and the property of symmetric function $p(x|\theta)=p(-x|\theta)$, the FIM can be computed to be (ds)^{2}=\sigma^{-1} (A(d\mu)^{2}+B(d\sigma)^{2}) which implies a constant negative scalar curvature. 
In fact, there is a deeper connection between the Gaussian and Cauchy distributions: they both belong to the location scale family distributions 
$f(x)=c^{-1}f((x-\delta)/c)$ where $\delta$ is the location
parameter, and $c$ is the scale parameter.  It is known any location scale distribution
has a constant negative curvature \cite{kass2011geometrical}. That the Gaussian and Cauchy distributions share this property but are distinct indicates that locally isometric
is not enough to distinguish them. This demands the use of a global distance as an additional measure
to characterize the model manifold. 
We embed the Cauchy distribution manifold using the isKL embedding with the distance measure \cite{chyzak2019closed}, which gives
\begin{equation}
    D^{2}_{sKL}(x_{1},\gamma_{1},x_{2},\gamma_{2})=2\log\bigg(\frac{(\gamma_{1}+\gamma_{2})^{2}+(x_{1}-x_{2})^{2}}{4\gamma_{1}\gamma_{2}}\bigg)
\end{equation}
Interestingly, the isKL embedding returns an Euclidean embedding for the Cauchy manifold (Fig.~\ref{fig:cauchy}), to the number of components we have explored. To compare it with the Gaussian fits manifold, we have colored the Cauchy manifold with Escher's art --- Circle Limit IV as well. Here, we observe well preserved bat shapes as compared to Fig.~\ref{fig:gauss}. Strikingly, not only this is also true for any symmetrized R\'enyi choices as shown in Fig. \ref{fig:cauchy} (b), the projections obtained from different symmetrized R\'enyi choices 
% Not exactly the same, but close
appears to be virtually the same. Thus $D^{2}_{sKL}$ is not obviously better than other intensive R\'enyi divergences for models not in exponential families. 

\section{Summary}
\label{sec:Summary}

In this paper, we demonstrate that any $N$ parameter probabilistic model that
takes the form of an exponential family can be embedded isometrically into a
low dimensional  $(N+N)$ Minkowski space via the isKL embedding technique.
This is done by using the symmetrized Kullback-Leibler divergence (sKL) as the
pairwise distance between model predictions. This could potentially be used to determine the number of parameters needed to describe an experiment or a simulation should the underlying distribution belongs to the exponential family. To illustrate how the isKL
embedding technique can be used to visualize the exponential family
probabilistic manifold in a simple and tractable way,  we consider the Bernoulli (coin
toss) problem, the ideal gas, the $n$ sided die, the nonlinear least square
models,  Gaussian fits to data, and the two dimensional Ising model.
Additionally, we use the non-exponential Cauchy distribution to illustrate
the importance of preserving both global and local structures in embeddings.
\appendix
\section{Replica Zero Limit of \textit{f} Divergence}
To visualize the underlying geometry of probabilistic model data, a distance measure in probability space is needed. In this appendix, we will generalize the limit of zero data procedure  in obtaining an intensive distance measure to a family of divergences, specifically from $f$ divergence to \Renyi\ divergence. $f$ divergence measures the difference between two probability distribution $P$ and $Q$  with a convex function $f$ such that $f(1)=0$ and takes the form 
\begin{equation}
D_{f}(P,Q)=\int f\bigg(\frac{p(x)}{q(x)}\bigg)q(x)d\mu(x)
\end{equation}
By assuming $f$ is analytic \cite{nielsen2013chi}, we can Taylor expand it about $x=1$, $f(x)=\sum_{m=0}^{\infty}\frac{1}{m!}f^{(m)}(1)(x-1)^{m}$. Thus, $f$ divergence takes the form
\begin{equation}
\begin{split}
D_{f}(P,Q)&=\int f\bigg(\frac{p(x)}{q(x)}\bigg)q(x)dx\\
&=\sum_{m=0}^{\infty}\int \frac{1}{m!}f^{(m)}(1)\bigg(\frac{p(x)}{q(x)}-1\bigg)^{m}q(x)dx\\
&=\sum_{m=0}^{\infty}\frac{1}{m!}f^{(m)}(1)\chi^{m}_{1,q}(P,Q)
\end{split}
\end{equation}
where 
\begin{equation}
\chi^{m}_{1,q}(P,Q)=\int\frac{(p(x)- q(x))^{m}}{q^{m-1}(x)}dx
\end{equation}
is the $\chi^{k}$-divergence with parameter $1$ . Expanding the polynomial and simplifying,

\begin{equation}
\begin{split}
 \chi^{m}_{1,q}(P,Q)&=\int\sum_{k=0}^{m} \genfrac(){0pt}{0}{m}{k}(-1)^{m-k}q^{1-k}(x)p^{k}(x)dx\\
&=\sum_{k=0}^{m}\genfrac(){0pt}{0}{m}{k}(-1)^{m-k}\int q^{1-k}(x)p^{k}(x)dx 
\end{split}
\end{equation}

Suppose we increase the number of data sample by $N$  which amounts to having $N$-replicated system,
\begin{widetext}
\begin{equation}
\begin{split}
\chi^{m}_{1,q}(P_{N},Q_{N})&=\sum_{k=0}^{m}\genfrac(){0pt}{0}{m}{k} (-1)^{m-k}\bigg(\int\ldots\int q^{1-k}(x_{1},..x_{N})p^{k}(x_{1},...,x_{N})dx_{1}\ldots dx_{N}\bigg)\\
&\bigg|\text{ Since } p(x_{1},...,x_{N})=\prod_{i=1}^{N} p(x_{i}) \text{ and  }  q(x_{1},...,x_{N})=\prod_{i=1}^{N} q(x_{i})\\
&=\sum_{k=0}^{m}\genfrac(){0pt}{0}{m}{k}(-1)^{m-k}\bigg(\int q^{1-k}(x)p^{k}(x)dx\bigg)^{N} \\
&=\sum_{k=0}^{m}\genfrac(){0pt}{0}{m}{k}(-1)^{m-k}\bigg[\bigg(\int q^{1-k}(x)p^{k}(x)dx\bigg)^{N}-1\bigg]+\sum_{k=0}^{m}\genfrac(){0pt}{0}{m}{k}(-1)^{m-k}\\
&\bigg|\text{ Note that } (1-x)^{n}=\sum_{n=0}^{\infty}\genfrac(){0pt}{0}{n}{k}(-x)^{n}, \text{ so} \sum_{k=0}^{m}\genfrac(){0pt}{0}{m}{k}(-1)^{m-k}=0.\\
&=\sum_{k=0}^{m}\genfrac(){0pt}{0}{m}{k}(-1)^{m-k}\bigg[\bigg(\int q^{1-k}(x)p^{k}(x)dx\bigg)^{N}-1\bigg]
\end{split}
\end{equation}
\end{widetext}
Upon closer inspection, each $\chi^{m}$ term contains partition function like terms $\big(\int q^{1-k}p^{k} dx\big)^{N}$ that is known as Hellinger divergence of order $k$ that increase geometrically with $N$.  Upon sending $N$ continuously to zero, we have 
%\begin{widetext}
\begin{equation}
\begin{split}
&\lim_{N\rightarrow 0}\frac{\chi^{m}_{1,q}(P_{N},Q_{N})}{N}\\&=\sum_{k=0}^{m}\genfrac(){0pt}{0}{m}{k}(-1)^{m-k}\log\bigg(\int q^{1-k}(x)p^{k}(x)dx\bigg)
\end{split}
\end{equation}
%\end{widetext}
As $ D_{\alpha}(P,Q)=\frac{1}{\alpha-1}\log\big(\int p^{\alpha}q^{1-\alpha}dx\big)$ is the \Renyi\ divergence, 
\begin{equation}
\begin{split}
     \lim_{N\rightarrow 0}\frac{\chi^{m}_{1,q}(P_{N},Q_{N})}{N}=\sum_{k=0}^{m}\genfrac(){0pt}{0}{m}{k}(-1)^{m-k}(k-1)D_{k}(P,Q)
\end{split}   
\end{equation}
Thus for any $f$ divergences,
%\begin{widetext}
\begin{equation}
\begin{split}
&\lim_{N\rightarrow0}\frac{D_{f}(P_{N},Q_{N})}{N}\\&=\sum_{m=1}^{\infty}\sum_{k=0}^{m}\frac{f^{(m)}(1)}{m!}\genfrac(){0pt}{0}{m}{k}(-1)^{m-k}(k-1)D_{k}(P,Q)
\end{split}
\end{equation}
%\end{widetext}
\newpage

\section{Connections to Principal Component Analysis (PCA) and Multidimensional Scaling (MDS)}
\label{sec:connection}
The interested reader will note a connection to both Principal Component Analysis (PCA)~\cite{hotelling1933analysis} and Multidimentional Scaling (MDS)~\cite{torgerson1952multidimensional}.  Principal
component analysis uses the isometries of Euclidean space to optimally display
data in a space of many dimensions.  PCA translates the data to center it,
then uses singular value decomposition to rotate and diagonalize the `moment of
inertia' tensor of the data set. The data remains many dimensional, but PCA
allows one to examine the directions for which the data varies the most. The
principal components are the orthogonal directions which best describe the data
set -- minimizing the sum of squared distances of the remaining data from an
approximation restricted to the subspace they span. 

Multidimensional scaling generalizes these ideas to situations where the data
vectors are not known, but some measure of the pairwise distance is available.
MDS generates an isometric embedding maintaining the pairwise distances,
usually in a vector space of dimension equal to the number of data points.
Again, this manifold can rotate or translate for a given system depending on
the sampling used. Indeed, the eigensystem solved in MDS often has negative
eigenvalues~\cite{harol2006augmented, pkekalska2006non,pkekalska2004not}
corresponding to time-like coordinates, and changing the sampling can also
induce Lorentz boosts.  MDS, using the symmetrized Kullback-Leibler divergence
$D^{2}_{sKL}$ as the pairwise distance, in fact produces an \isKL\ embedding
\footnote{Also, the \inPCA\ embedding~\cite{QuinnCdBNS19} is precisely MDS
applied to the Bhattacharyya distance $\dsBhat$.}.  Our main result
(Eq.~\eqref{eq:symKLcoords}) implies that MDS applied with $D^{2}_{sKL}$ to
high-dimensional data produced by an $N$-parameter exponential family will
embed its predictions in a much smaller space, with only $N$ space-like and $N$
time-like non-zero coordinates.  Furthermore, the resulting manifold will be
given by the explicit \isKL\ embedding of Eq.~\ref{eq:symKLcoords} up to isometries.

We can now establish a connection with the Multidimensional Scaling (MDS) technique. Given $n$ sampled points from the
parameter space, MDS generates an embedding whose $i$th projection is given by
$\sqrt{\Lambda_{i}}v_{i}$, where $\Lambda_{i}$ and $v_{i}$ are the eigenvalue
and eigenvector of the double mean centered pairwise distance matrix,
$D_{c}^{2}=-(1/2)PD^{2}P$, where $P_{i,j}=1/n-\delta_{i,j}$ and $D^{2}$ is the
pairwise distance matrix.  Writing out the matrix explicitly, we have $(D^{2}_{c})_{i,j}=-\frac{1}{2}\big(D^{2}_{i,j}+\frac{1}{n^{2}}\sum_{k,k'}D^{2}_{k,k'}-\frac{1}{n}\sum_{k}(D^{2}_{i,k}+D^{2}_{k,j})\big)$. We will solve for the eigensolutions in a more general setting by taking a continuous sampling limit. This yields an integral eigenvalue problem, 
\begin{equation}
    \int D^{2}_{c}(\btheta,\bgamma)v(\btheta)d\mu(\btheta)=\Lambda v(\bgamma),
    \label{eq:eign_ansatz}
\end{equation}
with
\begin{widetext}
\begin{equation}
    \begin{split}
        D_{c}^{2}(\btheta,\bgamma)&=-\frac{1}{2}\left( D^{2}(\btheta,\bgamma)
        -\int D^{2}(\btheta,\boldsymbol{\xi})d\mu(\boldsymbol{\xi})
        -\int D^{2}(\boldsymbol{\xi},\bgamma)d\mu(\boldsymbol{\xi})
        +\iint D^{2}(\btheta,\bgamma) d\mu(\btheta)d\mu(\bgamma) \right)
    \end{split}
\end{equation}

\end{widetext}
where $d\mu(\btheta)$ is the sampling measure, $v$ is the eigenfuntion and $\Lambda$ is the eigenvalue. One can recover MDS by having a discrete measure $\mu(\btheta)=\sum_{x}c_{x}\delta_{x}(\btheta)d\btheta$, where $\delta_{x}(\btheta)$ is the Dirac measure.  For $D^{2}_{sKL}$, the double mean centered distance measure takes the form
\begin{widetext}
\begin{equation}
        D^{2}_{c}(\btheta,\bgamma)=\frac{1}{2}\sum_{i}\bigg(\left(\eta_{i}(\bgamma)-\overline{\eta}_{i}\right)\left(\langle\Phi_{i}\rangle_{\btheta}-\overline{\langle\Phi_{i}\rangle}\right)
        +\left(\eta_{i}(\btheta)-\overline{\eta}_{i}\right)\left(\langle\Phi_{i}\rangle_{\bgamma}-\overline{\langle\Phi_{i}\rangle}\right) \bigg)
\end{equation}
\end{widetext}

where $\int \eta_{i}(\btheta)d\mu(\btheta)=\overline{\eta}_{i}$ and $\int \langle \Phi_{i}\rangle_{\btheta} d\mu(\btheta)=\overline{\langle\Phi_{i}\rangle}$. It turns out the coordinates $\mathcal{S}_{i}$ and $\mathcal{T}_{i}$ discussed in Sec. \ref{sec:isKLFamilies}
\begin{equation}
    \begin{split}
        \mathcal{S}_{i}(\btheta)&=\frac{1}{2}\left(\lambda_{i}\left(\eta_{i}(\btheta)-\overline{\eta_{i}}\right)+\frac{1}{\lambda_{i}}\left(\langle\Phi_{i}\rangle_{\btheta}-\overline{\langle\Phi_{i}\rangle}\right)\right)\\
        \mathcal{T}_{i}(\btheta)&=\frac{1}{2}\left(\lambda_{i}\left(\eta_{i}(\btheta)-\overline{\eta_{i}}\right)-\frac{1}{\lambda_{i}}\left(\langle\Phi_{i}\rangle_{\btheta}-\overline{\langle\Phi_{i}\rangle}\right)\right)
    \end{split}
\end{equation}
where $\lambda_{i}^{2}=\sqrt{\text{Var}(\langle\Phi_{i}\rangle)/\text{Var}(\eta_{i})}$
are indeed the solutions to Eq.~\eqref{eq:eign_ansatz}, with the $i$th eigenvalue pairs being $\Lambda^{\pm}_{i}=\frac{1}{2}(\text{Cov}(\eta_{i},\langle\Phi_{i}\rangle)\pm\sqrt{\text{Var}(\eta_{i})\text{Var}(\langle\Phi_{i}\rangle)})$. Here we will prove it as follows
\begin{widetext}
\begin{equation}
    \begin{split}
        &\int D_{c}(\btheta.\bgamma)\frac{1}{2}\left(\lambda_{i}\left(\eta_{i}(\btheta)-\overline{\eta}_{i}\right)\pm\frac{1}{\lambda_{i}}\left(\phib_{\btheta}-\overline{\langle\Phi_{i}\rangle}\right)\right)d\mu(\btheta)\\
        &\bigg| \text{ Letting }\int \eta_{i}(\btheta)\phib_{\btheta} d\mu(\btheta)=\overline{\langle\Phi_{i}\rangle\eta_{i}},\quad \int \eta^{2}_{i}(\btheta) d\mu(\btheta)=\overline{\eta^{2}_{i}} \text{ and } \int \phib^{2}_{\btheta} d\mu(\btheta)=\overline{\langle\Phi^{2}_{i}\rangle}\\
         &=\frac{1}{4}\left(\lambda_{i} \left(\eta_{i}(\bgamma)-\overline{\eta_{i}})\right)\left(\overline{\langle\Phi_{i}\rangle\eta_{i}}-\overline{\langle\Phi_{i}\rangle}\cdot\overline{\eta_{i}}\right)+\lambda_{i}\left(\langle\Phi_{i}\rangle_{\bgamma}-\overline{\langle\Phi_{i}\rangle}\right)\left(\overline{\eta_{i}^{2}}-\overline{\eta_{i}}^{2}\right)\right)\\
        &\quad \pm\frac{1}{4}\left(\frac{1}{\lambda_{i}}\left(\eta_{i}(\bgamma)-\overline{\eta_{i}})\right)\left(\overline{\langle\Phi^{2}_{i}\rangle}-\overline{\langle\Phi_{i}\rangle}^{2}\right)+\frac{1}{\lambda_{i}}\left(\langle\Phi_{i}\rangle_{\bgamma}-\overline{\langle\Phi_{i}\rangle}\right)\left(\overline{\langle\Phi_{i}\rangle\eta_{i}}-\overline{\langle\Phi_{i}\rangle}\cdot\overline{\eta_{i}}\right)\right)\\
        &\bigg| \text{ Rewriting }\overline{\langle\Phi_{i}\rangle\eta_{i}}-\overline{\langle\Phi_{i}\rangle}\cdot\overline{\eta_{i}}=\text{Cov}(\eta_{i},\langle\Phi_{i}\rangle), \quad \overline{\eta_{i}^{2}}-\overline{\eta_{i}}^{2}=\text{Var}(\eta_{i})\text{ and } \overline{\langle\Phi_{i}^{2}\rangle}-\overline{\langle\Phi_{i}\rangle}^{2}=\text{Var}(\langle\Phi_{i}\rangle)\\
        &=\frac{1}{4}\left(\text{Cov}(\eta_{i},\langle\Phi_{i}\rangle)\pm\frac{1}{\lambda_{i}^{2}}\text{Var}(\langle\Phi_{i}\rangle)\right)\lambda_{i}\left(\eta_{i}(\bgamma)-\overline{\eta}\right)\pm\frac{1}{4}\left(\text{Cov}(\eta_{i},\langle\Phi_{i}\rangle)\pm\lambda_{i}^{2}\text{Var}(\eta_{i})\right)\frac{1}{\lambda_{i}}\left(\langle\Phi_{i}\rangle_{\bgamma}-\overline{\langle\Phi_{i}\rangle}\right)\\
        &\bigg|\text{ Since } \lambda_{i}^{2}=\sqrt{\text{Var}(\langle\Phi_{i}\rangle)/\text{Var}(\eta_{i})} \\
        &=\frac{1}{2}\left(\text{Cov}(\eta_{i},\langle\Phi_{i}\rangle)\pm\sqrt{\text{Var}(\eta_{i})\text{Var}(\langle\Phi_{i}\rangle)}\right)\frac{1}{2}\left(\lambda_{i}\left(\eta_{i}(\bgamma)-\overline{\eta_{i}}\right)\pm\frac{1}{\lambda_{i}}\left(\langle\Phi_{i}\rangle_{\bgamma}-\overline{\langle\Phi_{i}\rangle}\right)\right)\\
    \end{split}
\end{equation}
\end{widetext}
As promised, $\mathcal{S}_{i}(\btheta)$ and  $\mathcal{T}_{i}(\btheta)$ are indeed the solutions to Eq.~\eqref{eq:eign_ansatz} with eigenvalues \begin{equation}
    \Lambda_{i}^{\pm}=\frac{1}{2}\left(\text{Cov}(\eta_{i},\langle\Phi_{i}\rangle)\pm\sqrt{\text{Var}(\eta_{i})\text{Var}(\langle\Phi_{i}\rangle)}\right).
\end{equation} 
In general, when the eigenvalues are degenerate, the axis of projections are free to rotate within the degenerate spacelike and timelike
subspaces, depending on $d\mu$. Hence, the solution  will be a linear
combination of the degenerate coordinates described in
Eq.~\eqref{eq:eign_ansatz}, i.e. 
$\mathcal{S'}(\btheta)=\sum_{k}\alpha_{k}\mathcal{S}_{k}(\btheta)$ and $\mathcal{T'}(\btheta)=\sum_{k}\beta_{k}\mathcal{T}_{k}(\btheta)$ where
$\sum_{k}\alpha^{2}_{k}=1$ and $\sum_{k}\beta^{2}_{k}=1$  and the index $k$ runs over coordinates that share the same eigenvalue.  In all our examples
except the generalized die, symmetry keeps rotations from mixing directions and
the projection coordinates can be calculated from Eq.~\eqref{eq:eign_ansatz} regardless of degeneracy.  

\begin{figure*}[ht]
    \centering
    \includegraphics[scale=0.35]{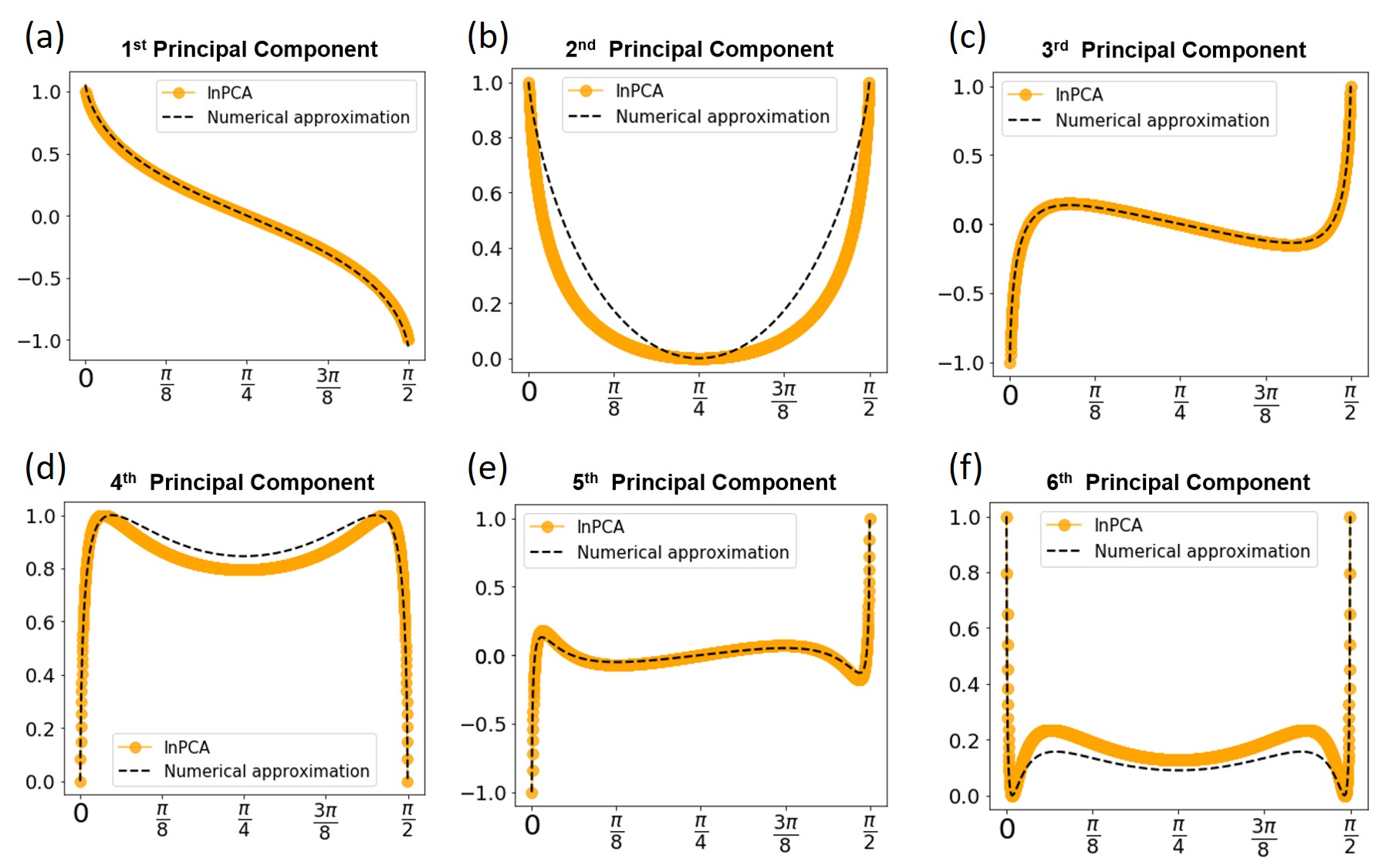}
    \caption{A-F. Normalized projection of coin toss manifold onto the first 6 principal axes. The dashed line is the numerical approximation of the analytical expressions given in Eq. B6 and Eq. B7 with $N=2000$} 
    \label{fig:appendix}
\end{figure*}
\section{Coin Toss and inPCA: The Bernoulli Problem model manifold embedded with the Bhattacharyya distance}
\label{sec:coin_sol}

In the Bernoulli problem, the inPCA embedding is given by the following pairwise distance
\begin{equation}
    d^{2}(\theta_{1},\theta_{2})=\log(\cos(\theta_{1}-\theta_{2}))
    \label{eq:coin}
\end{equation}
To find the embedding, we need to solve the eigenvalue problem discussed in Sec. \ref{sec:connection}. As the double mean centering matrix $P$ gives rotation and boost transformation to the coordinatess, for simplicity we proceed our calculation for each projection with just our distance function as an infinite matrix, acting on continuous variables $\phi$ and $\theta$: $\log \cos(\phi-\theta)$. This implies the evaluation of the following eigenvalue problem:
\begin{equation}
    \int_{0}^{\pi/2} \log\cos(\phi-\theta)v_{\alpha}(\theta)d\theta=\lambda_{\alpha}v_{\alpha}(\phi)
\end{equation}
where $v_{\alpha}(\phi)$ are the eigenfunctions with the coresponding eigenvalues $\lambda_{\alpha}$. We solve this numerically by expanding the pairwise distance function in terms of Chebyshev polynomials: $d^{2}(\theta,\phi)=-\log(2)+\sum_{k=1}^{\infty}\frac{(-1)^{k+1}}{k}\cos(2k(\theta-\phi))$ and assuming that the eigenfunction $v_{\alpha}(\theta)$ is odd with respect to $\theta=\pi/4$ and can be expanded
as Fourier series: $\sum_{k=1}^{\infty}b_{k}\sin(k(\theta-\frac{\pi}{4}))$. Thus we have
\par
\begin{equation}
\begin{split}
    \sum_{k,m=1}^{\infty}(-1)^{k+1}\frac{b_{m}}{k}F(\phi)=\lambda_{\alpha}\sum_{k=1}^{\infty}b_{k}\sin(k(\theta-\frac{\pi}{4}))
\end{split}
\end{equation}
with $F(\phi)=\int_{0}^{\pi/2} d\theta \cos(2k(\theta-\phi))\sin(m(\theta-\frac{\pi}{4}))$, where  As $F(\phi)$ only produces terms containing $\sin(2k(\phi-\frac{\pi}{4}))$ and  $\cos(2k(\phi-\frac{\pi}{4}))$ for all values of $m\in\mathbb{Z}^{+}$, it is thus natural to conjecture that the Fourier series expansion must have its coefficient $b_{2k+1}=0$. Hence,
\begin{equation}
    v_{\alpha}(\theta)=\sum_{k=1}^{\infty}b_{2k}\sin(2k(\theta-\frac{\pi}{4}))
\end{equation}
With this assumption, the eigenvalue equation simplifies into matching the coefficient of each Fourier mode $\sin(2k(\phi-\pi/4))$:
\begin{equation}
    \sum_{m=1}^{\infty}\xi(k,m)b_{2m}=\lambda_{\alpha}b_{2k}
\end{equation}
or more succinctly, $\xi\vec{b}=\lambda_{\alpha}\vec{b}$ where $\vec{b} = (b_{2}, b_{4},..., b_{2N}, ...)$. The matrix $\xi(k,m)$ is computed via $F(\phi)$ to be 
\begin{widetext}
\begin{equation}
    \xi(k,m)=\begin{cases}
      \frac{(-1)^{k+1}}{k}\frac{\pi}{4} & (m=k) \\
      \frac{(-1)^{k+1}}{k}\frac{1}{m^{2}-k^{2}}(k\cos(\frac{k\pi}{2})\sin(\frac{m\pi}{2})-m\cos(\frac{m\pi}{2})\sin(\frac{k\pi}{2})) & (m\neq k) \\
\end{cases} 
\end{equation}
\end{widetext}

For even eigenfunctions $v_{\alpha}(\theta)=\sum_{k=0}^{\infty}c_{k}\cos(k(\theta-\pi/4))$, the argument is almost identical, except
we now have an extra contribution from the constant $c_{0}$ term which needs to be handled separately. Going through the same derivation, we again have the matrix eigenvalue equation, i.e. $\eta\vec{c}=\lambda_{\alpha}\vec{c}$, where $\vec{c}=(c_{0},c_{2},...,c_{2N})$ and we have 
\begin{widetext}
\begin{equation}
    \eta(k,n)=\begin{cases}
      -\frac{\pi}{2}\log(2) & (n=k=0) \\
      -\log(2)\sin(\frac{n\pi}{2}) & (k=0,n\geq 1) \\
      \frac{(-1)^{k+1}}{k^{2}}\sin(\frac{k\pi}{2}) & (k\geq1,n=0)\\
      \frac{(-1)^{k+1}}{k}\frac{\pi}{4} & (k=n\geq1)\\
      \frac{(-1)^{k+1}}{k}\frac{1}{n^{2}-k^{2}}(n\cos(\frac{k\pi}{2})\sin(\frac{n\pi}{2})-k\cos(\frac{n\pi}{2})\sin(\frac{k\pi}{2})) & (n\geq1,k\geq1,n\neq k) \\
      \end{cases}
\end{equation}
\end{widetext}
One could get numerical approximation for the analytical calculation above by taking $\eta$ and $\xi$ to be finite-dimensional matrix $N\times N$, where $N\gg 1$ as shown in Fig.~\ref{fig:appendix} .

\acknowledgments

We thank Pankaj Mehta and Anirvan Sengupta for suggesting the possible importance of MDS. H.K.T was supported by the Army Research Office through ARO W911NF-18-1-0032. H.K.T, J. K-D. , C.C.B and J.P.S. was supported by the National Science Foundation through grant NSF DMR-1719490. K.N.Q was supported by an NSERC FPGS-D Fellowship.
\par

%%%%%%%%%%%

\bibliography{SethnaRecs,citation}
%%%%%%%%%%%

\end{document}